\documentclass[manuscript,showpacs,amsmath,amssymb,9pt]{revtex4-1}
\pdfoutput=1
\usepackage{graphicx,anysize}
\usepackage{amsmath}
\usepackage{amssymb}
\usepackage{psfrag}
\usepackage{threeparttable}
\usepackage{float}
\usepackage{color}
\usepackage{varioref}
\usepackage{epic}
\usepackage{eepic}
\usepackage{makeidx}
\usepackage{epsfig}
\usepackage{supertabular}
\begin{document}
\title{Defining the Free-Energy Landscape of Curvature-Inducing Proteins on Membrane Bilayers}

\author{Richard W. Tourdot}
 \email{tourdot@seas.upenn.edu}
\affiliation{Department of Chemical and Biomolecular Engineering, University of Pennsylvania, Philadelphia, PA, 19104}
\author{N. Ramakrishnan}
 \email{ramn@seas.upenn.edu}
\affiliation{Department of Bioengineering, University of Pennsylvania, Philadelphia, PA, 19104}
\author{Ravi Radhakrishnan}
 \email{rradhak@seas.upenn.edu}
\affiliation{Corresponding Author; Department of Chemical and Biomolecular Engineering, Department of Bioengineering, University of Pennsylvania, Philadelphia, PA, 19104}

\date{\today}
\begin{abstract}
Curvature-sensing and curvature-remodeling proteins, such as Amphiphysin, Epsin, and Exo70, are known to reshape cell membranes, and this remodeling event is essential for key biophysical processes such as tubulation, exocytosis, and endocytosis. Curvature-inducing proteins can act as curvature sensors; they aggregate to membrane regions matching their intrinsic curvature; as well as induce curvature in cell membranes to stabilize emergent high curvature, non-spherical, structures such as tubules, discs, and caveolae.  A definitive understanding of the interplay between protein recruitment and migration, the evolution of membrane curvature, and membrane morphological transitions is emerging but remains incomplete. Here, within a continuum framework and using the machinery of Monte Carlo simulations, we introduce and compare three free-energy methods to delineate the free-energy landscape of curvature-inducing proteins on bilayer membranes. We demonstrate the utility of the Widom test-particle/field insertion methodology in computing the excess chemical potentials associated with curvature-inducing proteins on the membrane\textemdash in particular, we use this method to track the onset of morphological transitions in the membrane at elevated protein densities. We validate this approach by comparing the results from the Widom method with those of thermodynamic integration and Bennett acceptance ratio methods. Furthermore, the predictions from the Widom method have been tested against analytical calculations of the excess chemical potential at infinite dilution. Our results are useful in precisely quantifying the free-energy landscape, and also in determining the phase boundaries associated with curvature-induction, curvature-sensing, and morphological transitions.  This approach can be extended to studies exploring the role of thermal fluctuations and other external (control) variables, such as membrane excess area, in shaping curvature-mediated interactions on bilayer membranes.
\end{abstract}

\preprint{Published as Phys. Rev. E {\bf 90}, 022717 (2014) }

\maketitle

\section{Introduction} 
Membranes constitute the boundary of all cells and cell organelles; these structures are primarily composed of a lipid bilayer. The curvature of a membrane (i.e., the curvature of the lipid bilayer) is considered to play an active role in controlling the spatial inhomogeneity and functionality in cells. Several membrane bound proteins are thought to be involved in generating and regulating membrane curvature, while many others sense background membrane curvature generated through other means. The mechanisms of membrane curvature generation and sensing have been classified into several categories based on their distinct qualitative features \cite{Baumgart:2011en,McMahon:2005km}. They include (1) {\em protein scaffolding}: in this mechanism multiple proteins locally concentrate to a region of the membrane and induce curvature by virtue of an intrinsic curvature in their membrane facing domains \cite{Zimmerberg:2005jk}, (2) {\em hydrophobic insertion}: in this mechanism, the involved proteins insert their hydrophobic domains into the membrane bilayer (also known as wedging) to generate curvature \cite{McMahon:2005km}, and (3) {\em oligomerization}: certain proteins, which cannot induce or sense membrane curvature individually, associate into oligomeric domains and induce curvature cooperatively \cite{McMahon:2005km,Zimmerberg:2005jk}. Examples of curvature-inducing proteins include families of proteins with membrane adjacent BAR (Bin/Amphiphysin/RVs) domains \cite{Peter:2004eb,Zimmerberg:2004ed,Gallop:2006dm}. BAR domains are crescent-shaped $\alpha$-helical bundles that bind to the membrane bilayer mainly through the processes of protein scaffolding and hydrophobic insertion. Based on their detailed structures the BAR domains are further sub-classified into classical BAR, N-BAR, F-BAR, etc. Another example is the dynamin family of proteins, which are  comprised of PH domains. A third example corresponds to proteins that employ hydrophobic insertion mechanism to generate curvature. Typically, these proteins have an intrinsically-disordered structure. Upon binding to membrane they undergo a folding transition to form amphipathic $\alpha$-helices which are buried inside a leaflet of the bilayer\textemdash specific examples include ENTH and ANTH domain containing proteins \cite{Baumgart:2011en,McMahon:2005km,Zimmerberg:2005jk}. For further discussion, we refer to a recent review article on  mechanisms of curvature induction by proteins on a bilayer membrane \cite{Ramakrishnan:2014prep}.

Most theoretical studies on protein binding are concerned with the adsorption on planar lipid bilayers (reviewed in \cite{Baumgart:2011en}); these studies are mainly concerned with planar membranes and curvature effects were not discussed in these studies. Reynwar et al. \cite{Reynwar:2007km} performed coarse-grained molecular dynamics simulations which show that once adsorbed onto lipid bilayers, curvature-inducing proteins experience an effective curvature-mediated attractive interactions. Jiang and Powers \cite{Jiang:2008je} investigated lipid sorting induced by curvature for a binary lipid mixture using a phase-field model. Das and co-workers have investigated the effect of protein sorting on tubular membranes using theoretical techniques \cite{Singh:2012gb,Zhu:2012eg}. Using Monte Carlo simulations, Sunil Kumar and coworkers \cite{SunilKumar:1999bf,SunilKumar:2001cr,Ramakrishnan:2010hk,Ramakrishnan:2013gl,Ramakrishnan:2012dk} studied the effects of membrane shape transitions and protein-induced anisotropic bending elasticity and curvature have on the shape of vesicles and the distribution of proteins on them. Using the Monte Carlo method, Liu et al., and Ramanan et al. have investigated the spatial segregation, curvature-sensing, and vesiculation in bilayers with curvature-inducing proteins \cite{Liu:2012es,Ramanan:2011ds}. Along with these computational studies, several experimental studies have been carried out to investigate curvature generation.

Sorre and coworkers \cite{Sorre:2009do,Sorre:2012if} have conducted experimental investigations into the sorting of lipids on a lipid membrane tube (tether) drawn from a giant unilamellar vesicle (GUV) using an optical trap. Curvature sorting of lipids and its influence on the bending stiffness of the bilayer membrane was studied by Tian et al \cite{Tian:2009fu,Tian:2009hx}. Dynamic sorting of lipids and proteins has been studied by Heinrich and coworkers \cite{Heinrich:2010ej}. These authors observed that the nucleation of disordered membrane domains occurs at the junction between the tether and GUV. Several other theoretical and experimental studies have helped shed light on the phenomena of curvature-mediated sorting \cite{Julicher:1996co, Heinrich:2010ej,Seifert:1993wq,Capraro:2010jo}. 

In general, curvature-inducing proteins can act as curvature sensors and aggregate on the  curved regions of the membrane \cite{Ramakrishnan:2013gl}. In this way cells can perform protein and lipid sorting for subsequent functions. The composition of lipids in a membrane may also modulate the curvature-sensing and curvature-generation activities of the proteins. Regulation of membrane curvature and its sensing is also important in understanding the underlying cellular physiology governed by trafficking, especially in the context of health conditions in humans. A definitive understanding of the interplay between protein binding/migration and membrane curvature evolution is emerging but remains incomplete. The mechanisms that underpin such behavior are hugely important in intracellular assembly and stability of organelles (which often sustain extreme curvatures). These mechanisms are also important in intracellular transport, and sorting of proteins and cargo. Though many aspects of these fundamental processes are well-characterized from a molecular biology perspective, especially in the domain of protein-protein interactions and increasingly in the area of protein localization, several open questions remain unanswered.  These form the basis for a complete understanding of the underlying mechanisms in these fundamental (``unit'') cellular process from a biophysical and thermodynamic perspective. The emerging picture from a wide array of recent studies is that molecular interactions between the protein and the lipids at the molecular scale directly determine the morphology of cellular membranes at the micron scale primarily by setting up curvature fields \cite{Zhao:2013hi,Ayton:2010gv,Cui:2009gd,Cui:2013dx,Tourdot:2014iet}. Determination and characterization of these curvature fields is a challenging task \cite{Lai:2012hk,Voth:2013he,Simunovic:2013ho,Arkhipov:2009fi,Zhao:2013hi}.

In cell membranes, protein-induced radius of curvature ranges from a few nanometers to a few tens of nanometers depending on the protein and lipid composition of the membrane. For example, N-BAR domains stabilize curvature regions with radius of mean curvature of 6.25~nm~\cite{Mim:2012je}, while dynamin induced tubes have radius of $25~$nm \cite{Marino:2005gl}. {\em In vitro} experiments have reported epsin induced tubulation of lipsomes with average tubule radius of $10~$nm~\cite{Lai:2012hk}. How these dimensions are related to the curvature induced by just one functional unit (i.e., the minimal oligomer with ability to induce persistent curvature) is not known. Multiscale modeling studies have been recently carried out to shed light into this important question \cite{Tourdot:2014iet,Lai:2012hk,Zhao:2013hi}. The membrane mediated interactions between the different curvature induced regions can extend beyond the range of the molecular size of these proteins. Hence in order to account for the disparate length scales, the effect of multiple proteins, and thermal fluctuations we adopt a continuum approach in this article which is based on the Canham-Helfrich description of membranes \cite{Helfrich:1973td}. The subject of this article is to define and quantify the free-energy landscape of such curvature-inducing proteins on a fluid bilayer membrane within the context of membrane elasticity theory. This  manuscript does not model specific experimental systems but rather focusses on a methodology to compute free energies. It relies on the premise that for proteins that induce curvature, when they act in dilute concentrations, the curvature induction is localized because there is only a finite amount of binding free energy available to deform the membrane. Hence, the spontaneous curvature around the protein will be localized. At the continuum level, just like the Canham-Helfrich description of the membrane as an infinitesimally thin surface, irrespective of the mechanism of curvature induction, we make the assumption that the effect of the protein is to introduce a curvature field (defined as the function $H_0$, see below), which is our definition for spontaneous curvature.

\section{Model}

\subsection{Membrane Model}
\label{sec:membranemodel}
For biological membranes\textemdash if the thickness is negligible when compared to its lateral dimensions\textemdash the thermodynamic behavior of the membrane can be well captured by the elastic energy functional \cite{Helfrich:1973td}
\begin{equation}
\mathcal{H} = \int \left(\frac{\kappa}{2} \left(2H - H_0\right)^2 + \bar{\kappa} K + \sigma_{\rm bare}\right) dA, 
\label{eqn:Helf-continuum}
\end{equation} 
where the material properties are given by $\kappa$, the bending rigidity; $\bar{\kappa}$, the saddle splay modulus; and $\sigma_{\rm bare}$, the bare surface tension. The geometric properties of the surface are given by the gauge-invariant scalars $H$ and $K$, the mean and Gaussian curvatures, respectively. $H_0$ is a spontaneous curvature field that represents the curvature-inducing interactions between the protein and membrane; see section~\ref{sec:memb-prot-model} for details. The integral is performed over the surface area of the membrane with $dA$ being the differential area. This approach of treating the effect of the curvature-inducing protein as a curvature field in the continuum field formulation has been utilized in prior studies \cite{Weinstein:2006fd,Agrawal:2010iu,Agrawal:2008ff,Agrawal:2009bt,Liu:2012es, Zhao:2013hi, Ramakrishnan:2012dk,Ramakrishnan:2011cc,Ramakrishnan:2010hk,Ramakrishnan:2013gl}.

\begin{figure}[!h]
\centering
\includegraphics [width=7.5cm] {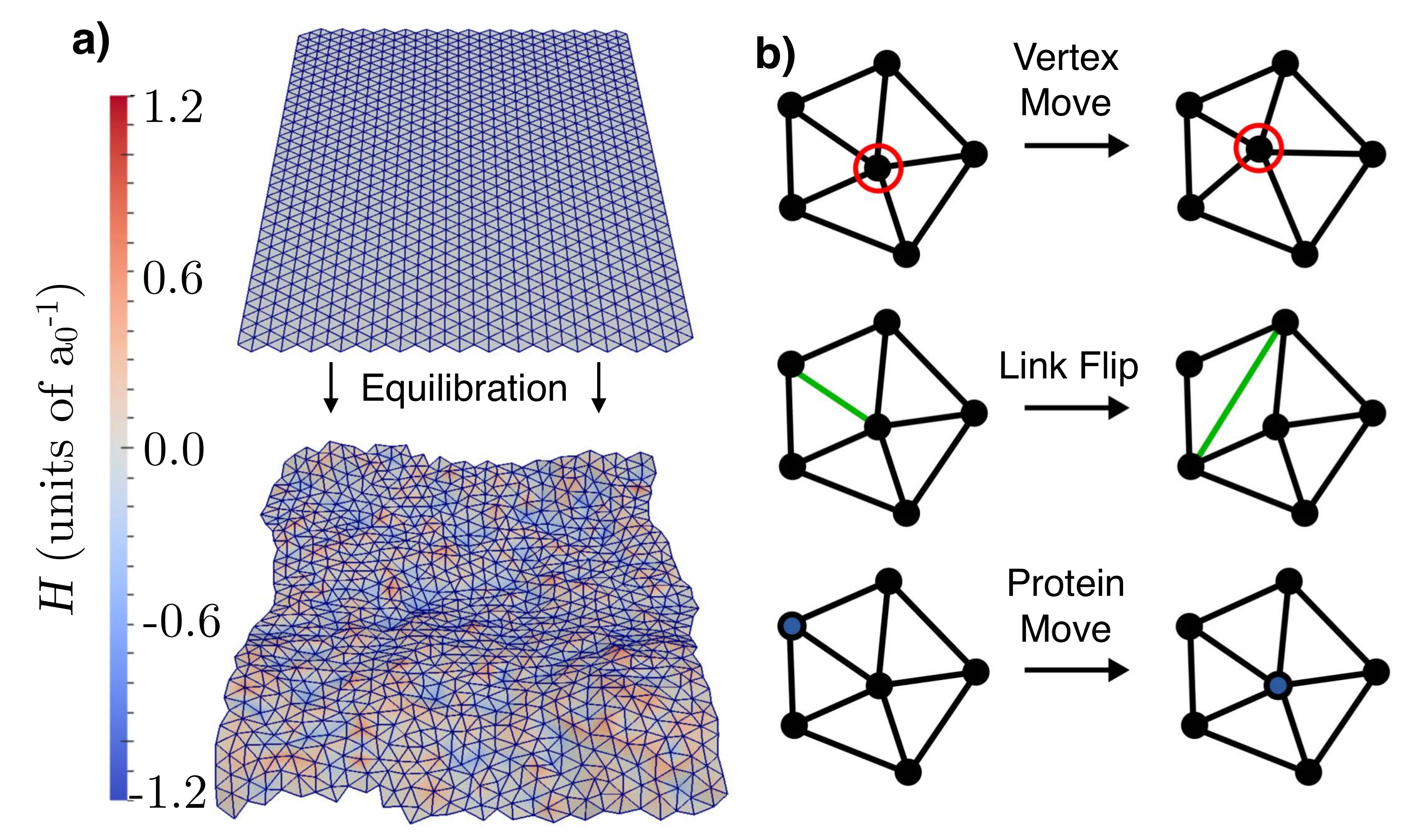}  
\caption{{\small \sl(Color Online)} a) Confomations of an initialy planar (top panel) and an equilibrated membrane (bottom panel); the vertices are colored based on the mean curvature which is expressed in units of $a_0^{-1}$.  b) The three Monte Carlo moves\textemdash namely, the vertex move, link flip, and protein move\textemdash used to evolve the membrane are shown.  \label{fig:memb-geometry}}  
\end{figure}

We make the system amenable to numerical simulations by discretizing the continuous membrane into a triangulated surface with $N$ vertices, $T$ triangles, and $L$ links. Self-avoidance is imposed by restricting the link length $l$ to be in the range $a_0\leq l \leq \sqrt{3}  a_0$. Here, $a_{0}$ is the characteristic length scale of the membrane which is much smaller the persistence length. We note that the length scale in the model is set by the value of $a_0$. We chose $N=900$ and initially place them on a square planar configuration as a $30 \times 30$ grid. The open edges of the membrane are subjected to periodic boundary conditions along the plane of the membrane (see Figure~\ref{fig:memb-geometry}). 

\subsection{Membrane-Protein Interaction Model} \label{sec:memb-prot-model}
For proteins that induce curvature, when they act in dilute concentrations, we expect their curvature induction to be localized because there is only a finite amount of binding free energy to deform the membrane. Hence, the spontaneous curvature around the protein will be localized to a finite length-scale. Irrespective of the mechanism of curvature induction, we make the assumption that the effect of the protein is to introduce a curvature field $H_0$. We justify our stance because the Canham-Helfrich formalism already approximates the model membrane to be infinitesimally thin. So the precise mechanism of spontaneous curvature induction, which can only be correctly modeled in an atomic level model, has to approximated in some manner within the Helfrich framework, which is through the choice of the $H_0$ function.  An alternative approach at the mesoscale is to explicitly represent the protein field as particles with suitably characterized membrane adhesion energies as done in \cite{Saric:2012hb,Bahrami:2012gb,Dasgupta:2013iz}. However we have chosen to employ the spontaneous curvature field model presented here since it fits well into a multi-scale framework, where the required field parameters can be determined from an all atom or coarse grained molecular simulation or experiment, as we describe below.

We do not know the exact nature of $H_0$. But several functions chosen for $H_0$ --- depending on the shape and extent of the function and depending on how many proteins are present --- will elicit a finite number of emergent membrane morphologies (such as vesicles, tubules, inward-tubules, caveloe etc.), a premise which is supported by a body of work discussed in reference \cite{Ramakrishnan:2014prep}.  For example, in earlier studies, we have shown that irrespective of whether we choose an isotropic Gaussian function or a cosine function or a square-well function, we will get vesicular buds under certain configurations \cite{Agrawal:2008ff,Liu:2012es,Agrawal:2010iu,Weinstein:2006fd}. Another example is that whether we choose an anisotropic (ellipse shaped) Gaussian dimple, or an anisotropic saddle shaped function, we can induce tubules \cite{Ramakrishnan:2014prep}. 

In this article, the spontaneous curvature induced in the vicinity of the membrane at $\vec{r}_{m}$ due to a protein at $\vec{r}_{p}$ is represented as,

\begin{equation}
H_0(\vec{r}_{m},\vec{r}_{p}) =  C_0\,{\cal F}(\vec{r}_{m},\vec{r}_{p}).
\end{equation}

Here $C_{0}$ is the induced membrane curvature at $\vec{r}_{m}=\vec{r}_{p}$. As a first approximation we choose this deformation profile ${\cal F}(\vec{r}_{m},\vec{r}_{p})$ to be a Gaussian function. A radially symmetric curvature profile has the form,

\begin{equation}
{\cal F}_{\rm iso}(r) = \exp \left(-\frac{r^{2}}{\epsilon^{2}}\right ),
\label{eqn:curiso}
\end{equation}
 
where $r=\left|\vec{r}_{m}-\vec{r}_{p}\right|$ and the $\epsilon^{2}/2$ is the variance of the Gaussian.  In general, the function ${\cal F}$ can take any arbitrary form as imposed by the protein curvature field. For instance, proteins containing BAR domains, like Endophilin and Exo70 domain containing Exocyst complex, are known to induce spatially anisotropic deformations \cite{Lai:2012hk,Voth:2013he,Simunovic:2013ho,Arkhipov:2009fi,Zhao:2013hi} that depends on the orientation of the protein $\theta=\arccos(\left \| \vec{r}_{m} \cdot \vec{r}_{p} \right \|)$. Such anisotropic curvature profiles can be modeled as,
 
\begin{equation}
{\cal F}_{\rm ani}(r,\theta)= \exp\left({-r^2 \left[\frac{\cos^{2}\theta}{\epsilon^{2}_{\parallel}} + \frac{\sin^{2}\theta}{\epsilon^{2}_{\perp}}\right]}\right).
\label{eqn:curaniso}
\end{equation}
 
$\epsilon^{2}_{\parallel}/2$ and $\epsilon^{2}_{\perp}/2$ are the variances along the directions parallel and perpendicular to the protein orientation, respectively.   

In order perform a systematic study of the free-energy landscape associated with curvature induction, we deal with curvature profiles that are analytically tractable: for this purpose we have chosen an isotropic spontaneous curvature profile in accordance with equation~\eqref{eqn:curiso}. Since this choice is an approximation to the exact shape of $H_0$, we discuss the question: for a given $H_0$ (e.g., equation~\eqref{eqn:curiso}), how do we estimate its parameters consistent with a given biological system?  

There are three methods we employ to determine the parameters of $H_0$ for a given biological system, which we summarize below. Method 1 (outlined in detail in previous work \cite{Agrawal:2010iu}) estimates the parameters in equation~\eqref{eqn:curiso} by matching the membrane deformation energy due to one spontaneous curvature field to the binding free energy of the protein with the membrane bilayer. Method 2 (outlined in reference~\cite{Liu:2012es}) estimates the parameters in equation~\eqref{eqn:curiso} by matching the computed curvature-induced sorting probability of the proteins with those measured in experiments. In Method~3, the numerical value of the field-parameters are determined based on molecular dynamics simulations at the atomic or near-atomic (coarse-grained) scales reported in the literature~\cite{Tourdot:2014iet,Zhao:2013hi}. In all three methods, the estimate for $C_0$ is $\sim 0.05$nm$^{-1}$ and that for $\epsilon$ is $\sim 17$~nm for ENTH domain proteins on a typical cell membrane with $\kappa=20 k_B T$. Later in the article, we set a typical value of $\epsilon^2= 6.3 {a_{0}}^{2}$, and $\kappa=10 k_B T$ (typical value for a lipid bilayer \textit{in vitro}), which fixes the value of $a_0 \sim 10$~nm.

On a triangulated membrane, though the core of each protein is defined on a vertex it induces curvature in the neighborhood of its core vertex in accordance with equation~\eqref{eqn:curiso}. Each of the $n$ proteins is associated with an unique vertex and each vertex can accommodate one protein at the most. The presence of multiple proteins in the vicinity of each other leads to a superposition of the spontaneous curvature fields. The exact form of  additivity of spontaneous curvature fields is not well established and hence we employ a simple additive rule where the multiple spontaneous curvature contributions at a given membrane location are linearly added and truncated as,
\begin{equation}
H_0(\vec{r}_m) = \min \left( 2C_{0},\underset{p=1}{\overset{n} {\sum}}H_0 (\vec{r}_{m},\vec{r}_{p}) \right).
\end{equation}
Note that $H_0(\vec{r}_m)$ denotes the total spontaneous curvature at membrane location $\vec{r}_m$ due to all proteins in its vicinity.

By including the effect of protein-membrane interaction as a spontaneous curvature field, we assume that the equilibrium behavior of the system is dominated by the membrane-mediated protein-protein interaction. These interactions are dictated by the strength and range of the curvature field and small-length-scale interactions (i.e., at the atomic level) are smoothed-out. Justification for this assumption has recently been presented by directly parameterizing such a curvature field from molecular dynamics simulations \cite{Zhao:2013hi}. In reference \cite{ArandaEspinoza:1996ek}, Aranda-Espinoza et al. employed a combination of integral equation theory and the linearized elastic free-energy model to describe the spatial distribution of the membrane-bound proteins. Their study indicates that the interaction (in the absence of thermal undulations) between two membrane-bound curvature-inducing proteins is dominated by a repulsive interaction. Consistent with these published reports, the calculated binding energy between two membrane-bound proteins   interacting through the curvature fields (again without thermal undulations) show dominant repulsive interactions which is governed by the range of the curvature field \cite{Agrawal:2008ff}.  Thus, purely based on energetic grounds, the previous analyses have suggested that membrane-deformation-mediated energies tend to be repulsive and should prevent, rather than promote, the formation of protein dimers or clusters.

Kozlov has discussed how the effect of fluctuations can change the repulsive nature of the interactions \cite{Kozlov:2007iaa}. The author's discussion is based on the premise that any membrane protein locally restrains thermal undulations of the lipid bilayer. Such undulations are favored entropically, and so this increases the overall free-energy of the bilayer. Neighboring proteins collaborate in restricting the membrane undulations and reduce the total free-energy costs, yielding an effective (membrane-mediated) protein-protein attraction. Indeed, for the linearized free-energy model, computing the second variation of energy, (note that at equilibrium, the first variation is zero, while the second variation governs the stiffness of the system against fluctuations) yields that the presence of a protein (or equivalently a curvature-inducing function) leads to a localized suppression of membrane fluctuations \cite{Agrawal:2008ff,Agrawal:2009bt}. This calculation has been further verified by using a free-energy method to compute the change in Helmholtz free-energy upon the introduction of a curvature field \cite{Agrawal:2009bt}. This provides for the possibility of  an entropically-mediated protein-protein attraction. The outcome of the interplay between the attractive entropic forces and the repulsive energetic forces is context specific as both have the same dependence on the protein-protein distance, and their absolute values differ only by coefficients with similar values. This has been demonstrated by examining the protein-protein pair correlation (spatial and bond-orientational) and through the effect on membrane morphology \cite{Agrawal:2008ff}. Indeed the model predicts that the cooperative effects of membrane-mediated interactions between multiple proteins can drive different morphological transitions in membranes \cite{Agrawal:2008ff,Reynwar:2007km,ArandaEspinoza:1996ek,Kozlov:2007iaa}. This notion of cooperativity is also consistent with the analysis of Kim et al. \cite{Chou:2001bm}, who have shown using an energetic analysis that in the zero temperature limit, clusters with size larger than five membrane-bound curvature-inducing proteins can be arranged in energetically stable configurations. It is also worth mentioning for completeness that Chou et al. \cite{Chou:2001bm} have extended the energetic analysis to membrane-bound proteins that have a noncircular cross-sectional shape and to local membrane deformations that are saddle shaped (negative Gaussian curvature) and have shown that in such cases the interactions can be attractive even without considering fluctuations.

\subsection{Monte Carlo Moves}
The accessible states of the membrane protein system are sampled using a set of three Monte Carlo moves that mimic membrane undulations, lipid diffusion, and protein diffusion. In the framework of Dynamically triangulated Monte Carlo (DTMC), a Monte Carlo step (MCS) comprises of $N$ attempts to randomly displace the vertices, $L$ attempts to flip the links and $n$ attempts to randomly displace the protein on the membrane surface. The various moves have been illustrated in Figure~\ref{fig:memb-geometry}b and each of the attempted moves are accepted using the Metropolis algorithm \cite{Frenkel:2001}.  For a complete description of the Monte Carlo moves see reference \cite{Liu:2012es}.

In our model, the random displacement vector is adaptively chosen to ensure that the acceptance rate for the vertex move is 50$\%$, while the acceptance rate for link flips and protein diffusion are dictated by the geometry.  All our simulations were equilibrated for 10 million MCS and statistics were collected over another 20 million MCS.


\subsection{Ensemble for the Planar Membrane}
\label{subsec:ensemble}
A planar membrane is characterized by its extensive variables; the entropy, $S$; the surface area, $A$; and the projected area, $A_{P}$. The internal energy of the membrane with $n$ proteins is given by,
\begin{widetext}
\begin{equation}
\begin{split}
dU(N,n,A,A_P,S)= d{\cal H} = \mu dN + \mu_{P}dn + \sigma dA+\gamma dA_{P}+TdS.
\end{split}
\label{eqn:internal-ener}
\end{equation}
\end{widetext}
Here the conjugate variables are $\mu$, the chemical potential of the membrane, $\mu_P$, the chemical potential of a membrane protein, and $\gamma$, the tension due to the frame (also called the frame tension). It should be noted that for closed membranes (e.g., a cylindrical membrane or a spherical vesicle) the volume enclosed by the membrane, $V$, and the osmotic pressure difference, $-P$, are used in place of $A_P$ and $\gamma$, respectively; we limit our studies in this article only to planar membranes. We can assume that $l$ is not a physically independent variable from the list of extensive variables defined above;  rather, $l$ sets the length-scale or resolution of the mesh, and tuning it allows us to change $A/A_{P}$ or the value of $\sigma$. Here, $\sigma$ is an effective tension conjugate to $A$ and is constituted by a combination of the bare surface tension ($\sigma_{\rm bare}$) and the area compressibility modulus. In our simulations, we control (hold constant) $N,n,\sigma,  A_{P}$, and $T$. Hence, the suitable thermodynamic potential for a planar membrane in our simulations is given by,
\begin{equation}
\begin{split}
dF(N,n,\sigma,A_{P},T)= d{\cal H} - TdS - SdT - \sigma dA - A d \sigma.
\end{split}
\label{eqn:pl-free-ener}
\end{equation} 
The effective surface tension $\sigma$ defined in equation~\eqref{eqn:internal-ener} should be distinguished from the bare surface tension $\sigma_{\rm bare}$ defined in equation~\eqref{eqn:Helf-continuum}. We have performed all our studies with $\sigma_{\rm bare}=0$. However, the effective surface tension determined from the fluctuation spectrum can still be non-zero, (see section \ref{sec:uspec}), because the value of $\sigma$ is renormalized by an effective area compressibility modulus term; the latter arises because of the constraint $a_0 \leq l \leq \sqrt{3} a_0$, we impose for self-avoidance, see section~\ref{sec:membranemodel}.

\section{free-energy Methods}
The free-energy landscape of the protein-membrane system drives key biophysical phenomena including protein recruitment, protein membrane remodeling, curvature-sensing, and protein clustering. Hence, in order to gain better insight into the behavior of this system, we delineate a strategy to compute the free-energy landscape for a single protein interacting with the membrane using the suite of free-energy methods described below.

\subsection{Widom Test-Particle/Field Insertion Method}
We determine the change in free-energy when a protein binds to the membrane by determining the excess chemical potential using the test-particle insertion method. The Widom particle/test-particle insertion method is a computational technique used to probe a system's chemical potential \cite{Widom:1963fl}. This technique samples the excess chemical potential by randomly inserting a virtual test (ghost) particle, and determines the change in the system's energy due to insertion of the test particle. 

Let $Q_{n}$ and $Q_{n+1}$ be the partition functions for a membrane with $n$ and $n+1$ proteins, respectively. The partition function is related to the configurational free-energy, (i.e., not including the contribution from the kinetic energy or from internal degrees of freedom such as rotation), as $F_{n}=-k_{B}T \ln Q_{n}$ for all $n$. Hence, the change in free-energy upon insertion of a protein field (i.e., the test particle) in a membrane with $n$ proteins is given by: 
\begin{equation}
\Delta F=F_{n+1}-F_{n}=-k_{B}T \ln\left(\frac{Q_{n+1}}{Q_{n}} \right).
\label{eqn:widom-freeener}
\end{equation} 
It can be seen from  equations~\eqref{eqn:internal-ener} and~\eqref{eqn:pl-free-ener} that the above change is equal to the chemical potential,
\begin{equation}
\mu_{P} = \left. \frac{\partial F}{\partial n} \right |_{A_P,\sigma,N,T}.
\label{eqn:mu}
\end{equation}
Combining equation~\eqref{eqn:mu} with equation~\eqref{eqn:widom-freeener} we obtain:
\begin{equation}
\mu_{P} = -k_B T \ln\left(\frac{Q_{n+1}}{Q_n}\right),
\end{equation}
 which can be decomposed into an ideal gas contribution and an excess contribution such that,
 \begin{equation}
  \mu_{P} = \mu_{P}^{\rm id} (\rho) + \mu_{P}^{\rm ex}.
  \label{eqn:widom}
  \end{equation} 
  The configurational component of the ideal part can be calculated from the protein density $\rho$ as $k_{B}T \ln \rho$; we note that the full ideal gas contribution is given by $\mu_P^{\rm id}(\rho) = k_B T \ln (\rho \Lambda^d)$ not including the contributions from the internal degrees of freedom.  Here, $\Lambda=(2 \pi m k_B T/h^2)^{-1/2}$ with $m$ the molecular mass of the protein, $h$ the Planck's constant, and $d$ the dimensionality of the system. If $\Delta {\cal{H}}$ be the energy change due to insertion of a test curvature-inducing protein then the excess chemical potential is written as,
\begin{equation}
\mu_{P}^{\rm ex} = -k_B T \ln{\int{\langle \exp(-\beta \Delta {\cal{H}}) \rangle P_{\rm uniform} (s_{n+1}) ds_{n+1}}}.
\label{eqn:muex}
\end{equation}
Here, $\beta=(k_B T)^{-1}$, and $\Delta {\cal{H}} = {\cal{H}}\left({n+1}\right) - {\cal{H}}\left(n\right)$ and the ensemble average $\langle  \cdot \rangle$ is taken over the phase space defined by the membrane and the $n$ protein fields. Here, $s_{n+1}=\vec{r}_{p}$, with $p=n+1$, is the position of the $n+1^{th}$ protein field, and $P_{\rm uniform} (s_{n+1})$ represents a uniform probability distribution from which the coordinate of the $n+1^{th}$ particle/field is sampled. The integral over $s_{n+1}$ amounts to the sum over all Widom test paticle/field insertion trials, and $P_{\rm uniform} (s_{n+1})$ equals the reciprocal of the total number of trials. For conciseness, we represent the right-hand-side term in equation~\eqref{eqn:muex} as $- k_B T\ln\langle \exp\left(-\beta\Delta {\cal{H}} \right) \rangle_{n}$. This formulation  is derived for a homogeneous membrane while the corresponding form of equation~\eqref{eqn:muex} for a spatially inhomogeneous membrane, where $\mu_{P}^{\rm ex}$ is a function of the phase space variables ${\bf r}$, is given by, 
\begin{equation}
\mu_{P}^{\rm ex}({\bf r}) = - k_B T\ln\langle \exp\left(-\beta\Delta {\cal{H}}({\bf r})\right) \rangle_{n}.
 \label{eqn:inhomo}
\end{equation}
At equilibrium the bulk chemical potential $\mu_{P}$ is a constant, hence the scaled, inhomogeneous, spatial density can be determined as,
\begin{equation}
\rho({\bf r}) = \rho_{0}\langle\exp\left(-\beta\Delta {\cal{H}}({\bf r})\right) \rangle_{n},
 \label{eqn:inhomo-density}
\end{equation}
where $\rho_{0} = \exp\left(\mu_{P}\right)$.
The Widom test-particle/field insertion method is more suitable to probe chemical potentials in dilute systems whereas its applicability to systems with large protein concentrations is limited; see Appendix \ref{app:widom} for a discussion. Hence, in order to study the higher protein concentrations we also use more reliable methods based on free-energy perturbation, which are defined in the next sections.

\subsection{Thermodynamic Integration (TI) Method}
Thermodynamic integration is a free-energy perturbation technique used to compute the change in free-energy  between two states $A$ and $B$, with energies ${\cal H}_{A}$ and ${\cal H}_{B}$; these states correspond to a membrane with $n$ and $n+1$ proteins, respectively. Further, state $A$ is characterized by a scalar parameter $\lambda=0$ and state $B$ by $\lambda=1$. The system is evolved with a Hamiltonian (or energy function) ${\cal H}(\lambda)=(1-\lambda){\cal H}_{A}+\lambda {\cal H}_{B}$. To define a path between $A$ and $B$, the parameter $\lambda$ is varied between $0\leq\lambda \leq 1$ in successive windows of the simulation. The free-energy change along this path \cite{Frenkel:2001} is given by,
\begin{equation}
\Delta F_{\rm TI} = F_{B} - F_{A} = \int_{0}^{1} \left\langle \frac{\partial {\cal H}(\lambda)}{\partial \lambda} \right\rangle d \lambda.
\label{eqn:TI}
\end{equation}
TI overcomes many of the limitations of the Widom test-particle/field insertion method (see Appendices~\ref{app:widom} and~\ref{app:ti}), but the results of equation~\eqref{eqn:TI} should match the results from equation~\eqref{eqn:widom} in the dilute limit; i.e., when the concentration of protein is such that $n<<N$.

\subsection{Bennett Acceptance Ratio Method (BAM)}
The Bennett acceptance method is also used to approximate the free-energy difference between two states close to each other in phase space.  This method is derived from the detailed balance equations involving two states ($A$ and $B$)  \cite{Bennett:1976gj}. Namely,
\begin{equation}
M({\cal{H}}_A - {\cal{H}}_B) \exp(-\beta {\cal{H}}_B) = M({\cal{H}}_B - {\cal{H}}_A) \exp(-\beta {\cal{H}}_A),
 \label{eqn:rates-equal}
\end{equation}
where $M$ is some function that defines the acceptance distribution for transition from state $A$ to state $B$ or vice versa. In our case, we choose $M$ to be the Metropolis function $M(x)=\min(1,\exp(- \beta x))$, which defines the acceptance probability according to a Boltzmann distribution. This yields:
\begin{equation}
\exp\left(\frac{- \Delta F_{\rm BAM}}{k_B T}\right)_{A\rightarrow B} = \frac{Q_B}{Q_A} = \frac{\left\langle M({\cal{H}}_B - {\cal{H}}_A) \right\rangle_A} {\left\langle M({\cal{H}}_A - {\cal{H}}_B) \right\rangle_B}.
\label{eqn:bennett-eqn}
\end{equation}
Appendix~\ref{app:bennett} provides a brief discussion of the expected accuracy of the Bennett acceptance methodology for the choice of the acceptance function $M$ described above; the Bennett acceptance method can be improved further by optimizing the function $M$, to decrease the sampling error \cite{deRuiter:2013iu}.

\subsection{Analytic Approximation to the Excess Chemical Potential of Curvature-Inducing Proteins}
For some special cases the chemical potential of a curvature-inducing protein can be derived analytically.  The change in energy due to the addition of one curvature protein can be determined from equation \eqref{eqn:Helf-continuum} as,
\begin{equation}
\Delta {\cal{H}}  = \int{\frac{\kappa}{2} \left(-4 H H_0 + H_0^2 \right)} dA.
\label{eqn:delU}
\end{equation}
At infinite dilution (i.e. when $n=0$) this change in energy for curvature fields given by equation~\eqref{eqn:curiso} can be included in the expression for the excess chemical potential, which can be expressed as:   

\begin{widetext}
\begin{equation}
\mu_{P}^{\rm ex} = -k_B T \ln{\left\langle \exp\left(\frac{- \kappa}{k_B T} \left(-2 C_0 \int H f(r) dA + \frac{\pi \epsilon^2 C_0^2}{4} \right)\right) \right\rangle}_{n=0}.
\label{eqn:analytic-excess}
\end{equation}

This relation can be further simplified to,  

\begin{equation}
\mu_{P}^{\rm ex}  =  \underbrace{\frac{\kappa \pi \epsilon^2 C_0^2}{4}}_{\mu_{T=0}} -\underbrace{k_B T \ln{\left\langle \exp\left(\frac{ 2 \kappa C_0 \int H f(r) dA }{k_B T}\right) \right\rangle}_{n=0}}_{\mu_{\rm fluc}},
\label{eqn:analytic-excess-simplified}
\end{equation}
\end{widetext}
since the second term in the exponential depends only on constants.  In the above equation, $\mu_{T=0}$ can be interpreted as the chemical potential to insert a protein on a flat membrane.  Cellular membranes can remain planar when the membrane is strongly bound or pinned to other cellular components like the cytoskeleton and other membrane binding proteins which can be characterized by a pinning fraction. The pinning fraction $\phi$ can range from 0 for an free membrane to 1 for a completely pinned membrane.  When $\phi < 1$ the chemical potential has additional contributions from the undulation modes of the membrane, which is given by $\mu_{\rm fluc}$.  

\section{Results}
\subsection{Membrane Undulations and Power Spectrum} \label{sec:uspec}
The equilibrium properties of an undulating membrane are significantly influenced by the choice of control variables (see section~\ref{subsec:ensemble}). Hence before delineating the protein induced deformations we first analyze the fluctuation modes of a planar membrane in the absence of a curvature (protein) field. The height-height correlation of a planar membrane, described by equation~\eqref{eqn:Helf-continuum}, parameterized  in the Monge Gauge \cite{Seifert:1997wq,Brown:2011iq}, and expressed in Fourier space, is given by,
\begin{equation}
\langle h_{q}h_{-q} \rangle =\frac{k_{B}T}{A_{P}\left[\kappa q^{4}+\sigma q^{2} \right ]}.
\label{eqn:hqhmq}
\end{equation}
Here, the angular brackets represent the equilibrium ensemble average, and we define $h_{q}$ as the 2-dimensional discrete Fourier transform of the membrane height function $h(\vec{r})=h(x,y)$. Namely,
\begin{equation}
h (\vec{r}) = \sum_{q} \, h_q \, \exp \left( i \vec{q} {\cdot} \vec{r} \right).
\label{eqn:h-fourier}
\end{equation}
In equation~\eqref{eqn:h-fourier}, $\vec{q}=(q_x, q_y)= 2 \pi (n_x/L,n_y/L)$, where $A_P=L^2$ and $n_x, n_y$ are integers.

\begin{figure}[!h] 
\centering
\includegraphics[width=7.5cm]{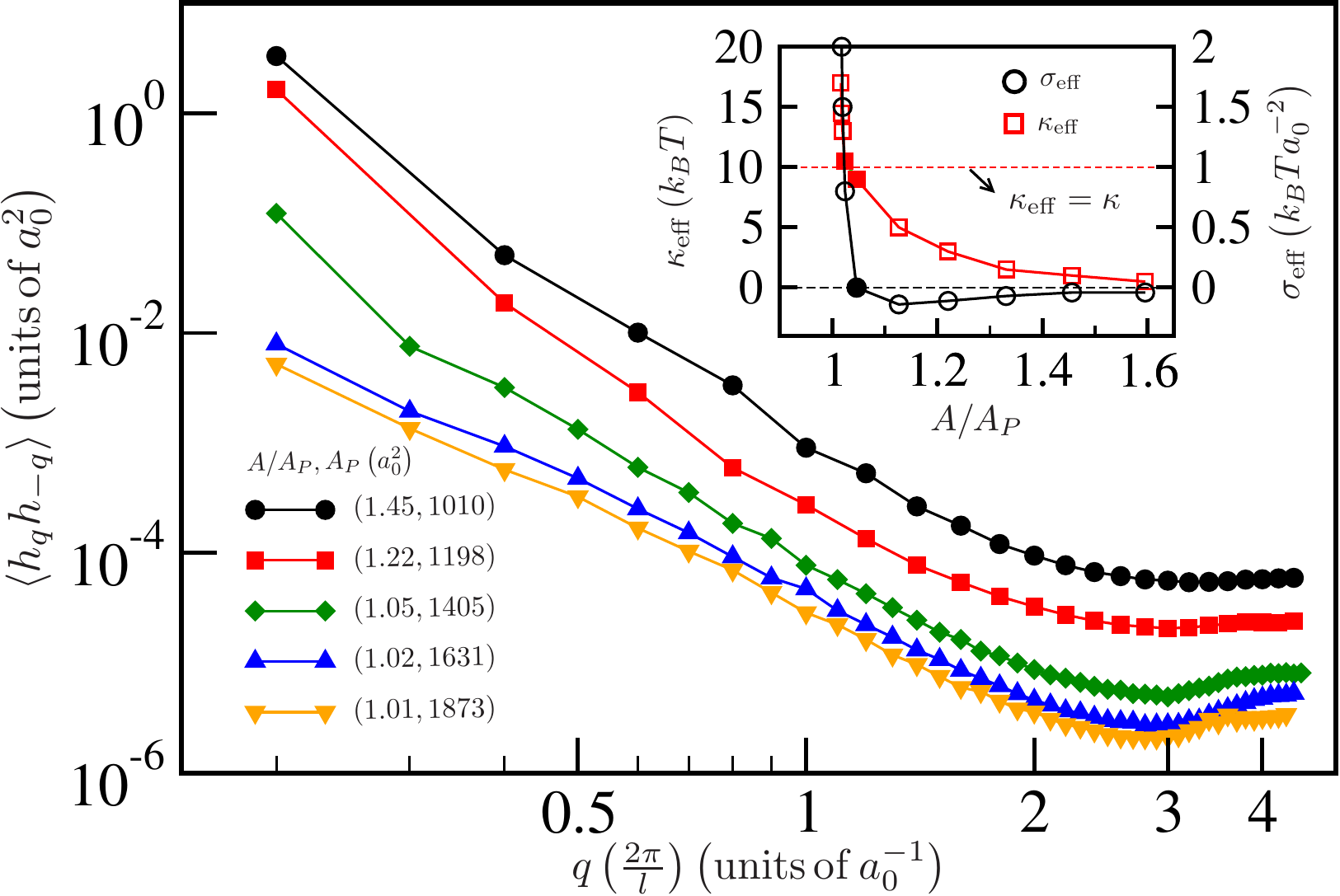}  
\caption{{\small \sl(Color Online)} Undulation spectrum (main plot) and the fit values for $\kappa_{\rm eff}$ and $\sigma_{\rm eff}$ (inset) for different values of $A/A_{P}$. Each pair of values seen in the legend of the main plot corresponds to $(A/A_P, A_P)$ for the membrane. With increase in $A/A_{P}$ the small $q$ behavior transitions from a concave to a convex profile, which is characteristic of $\sigma_{\rm eff}$ crossing over to negative values as shown in the inset. The effective bending rigidity is also renormalized with change in $A/A_{P}$ such that $\kappa_{\rm eff} \rightarrow 0$ as $A/A_{P} \rightarrow \infty$.  The filled symbols in the inset correspond to values of $A/A_{P}$ for which $\sigma_{\rm eff} \sim \sigma_{\rm bare}$ and $\kappa_{\rm eff} \sim \kappa$. \label{fig:freememb-char}}   
\end{figure}

The undulation spectrum corresponding to a planar membrane with $\kappa=10k_{B}T$ and $\sigma_{\rm bare}=0$ for a range of $A/A_{P}$ is shown in Figure~\ref{fig:freememb-char}. The data was fit to equation~\eqref{eqn:hqhmq} and the corresponding fit parameters, $\kappa_{\rm eff}$ and $\sigma_{\rm eff}$, are shown in the inset to Figure~\ref{fig:freememb-char}; also see Figure~S1 in the Supplementary Material.

When $A/A_{P}>1.05$ the membrane displays dominant long wavelength undulations, represented  in Figure~\ref{fig:freememb-char} by the large intensities of the power spectrum at low $q$; this results in shapes with curvatures of large magnitude. In this regime $\kappa_{\rm eff} <\kappa$\textemdash which corresponds to thermal (entropic) softening of the membrane\textemdash and $\sigma_{\rm eff} \sim 0$, which implies that the membrane is tensionless. Henceforth, we choose to model the membrane with $A/A_{P}=1.04$ (parameters corresponding to the filled symbols in the inset to Figure~\ref{fig:freememb-char}) for which we compute $\kappa_{\rm eff} \sim \kappa$ and $\sigma_{\rm eff} \sim \sigma_{\rm bare}+\sigma=0.0$; however, we note that $A/A_{P}$ is an important parameter which defines the thermodynamic ensemble in section~\ref{subsec:ensemble}. 

\subsection{Membrane Conformations versus $C_{0}$ and $n$}
\begin{figure}[!h]
\centering
\includegraphics[width=7.5cm,clip]{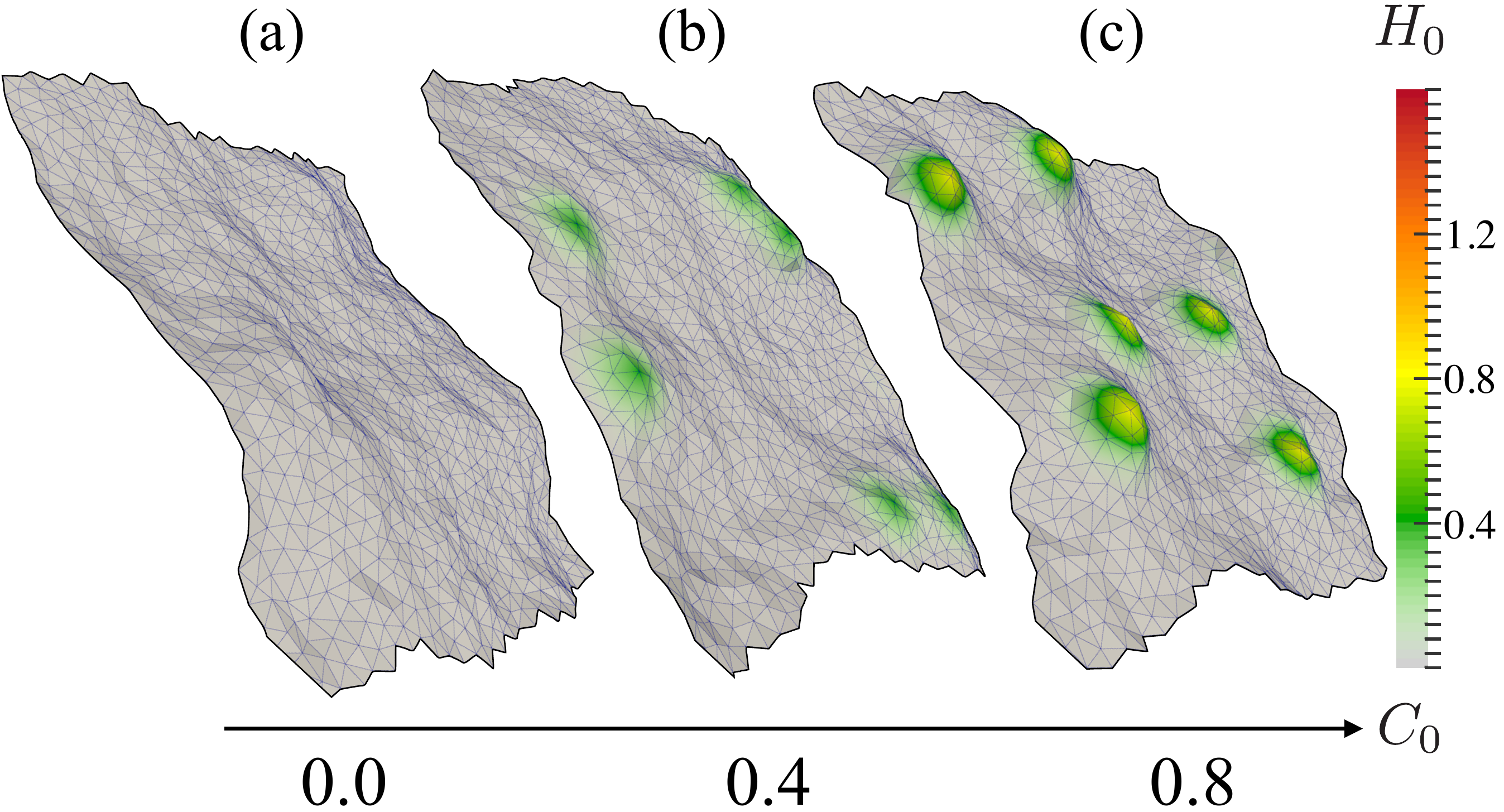}
\caption{\label{fig:epsin6-funcczero}{\small \sl(Color Online)} Representative membrane conformations as a function of imposed curvature $C_{0}$ for a system with 6 proteins: (a) no protein fields; (b) six protein fields each with $C_0=0.4 a_0^{-1}$; (c) six protein fields each with $C_0=0.8 a_0^{-1}$. Color bar shows the induced curvature field $H_{0}$ in units of $a_0^{-1}$.}
\end{figure}

The equilibrium shapes of a planar membrane interacting with spontaneous curvature-inducing proteins with fixed $\epsilon^{2}=6.3a_{0}^{2}$, and for different magnitudes of the imposed curvature $C_{0}$, is shown in Figure~\ref{fig:epsin6-funcczero}. 

A comparison of the membrane conformations for $C_{0}=0.0,\,0.4,\,$ and $0.8 {a_0}^{-1}$, in Figure~\ref{fig:epsin6-funcczero}, shows that in the presence of a small number of the curvature-inducing proteins (dilute limit) the membrane does not undergo any morphological changes, which is consistent with previous studies \cite{Agrawal:2008ff}.  This is characteristic of membranes with dilute protein concentrations or proteins imposing weak spontaneous curvatures. In the dilute limit the proteins localize to regions on the membrane matching their curvature field; however, the concentration is too low to promote any spatial aggregation of proteins which can lead to a morphological transition. Hence, the proteins in this concentration regime can largely be regarded as {\em curvature-sensors}. We note, however, that even in the dilute limit there is significant renormalization of the bending stiffness and membrane tension (see Figures ~S2 and ~S3 in the Supplementary Material). The effects at higher concentrations are more drastic leading to a change in the undulation behavior; that is, for larger $n$,  the corresponding governing equations are more complex than that described by equation~\eqref{eqn:hqhmq} as the undulation spectrum is two-dimensional and depends on $q, q'$. Specifically, when $\kappa= \kappa(x,y)$, $\sigma=\sigma(x,y)$, and $H_0 =H_0(x,y)$, and whose respective Fourier transforms are given by $\kappa_{q}, \sigma_{q}, h_{0,q}$, we can show that ${\cal H}$ is given by:
\begin{widetext}
\begin{equation}
\langle {\cal H} \rangle = {\frac{1}{2 A_P}} \, \sum_{\vec{q}} \sum_{\vec{q'}} \, \{  [  q^2 {q'}^2   \langle h_{q} h_{q'} \rangle -  {q}^2 \langle h_{q} h_{0,q'} \rangle  -  {q'}^2 \langle h_{0,q} h_{q'} \rangle +  \langle h_{0,q} h_{0,q'} \rangle  ]  \kappa_{q+q'} \, + \, q q' \, [ \langle h_{q} h_{q'} \rangle  ] \, \sigma_{q+q'} \}. 
\label{eqn:hqhmq-complex}
\end{equation}
\end{widetext}
At large $n$, as can be seen from Figures ~S2 and ~S3, the spontaneous curvature fields significantly influence the low $q$ modes of the fluctuation spectrum.

Our results in Figure ~S4 (Supplementary Material) also quantify the increase in excess area of the membrane $\left(A-A_{P}\right)$ as a function of $n$ for different values of $C_0$. We find the membrane excess area to increase with protein concentration ($n$), and is more pronounced for higher values of $C_0$. This increase is consistent with the softening of the membrane  (i.e., lowering of $\kappa$); however the effect is subtle because a positive renormalized tension is manifested. For larger $n$, the membrane becomes substantially softer, however, the undulation behavior is more complex than that described in equation~\eqref{eqn:hqhmq}, as discussed above.

\begin{figure}[!h]
\centering
\includegraphics[width=7.5cm,clip]{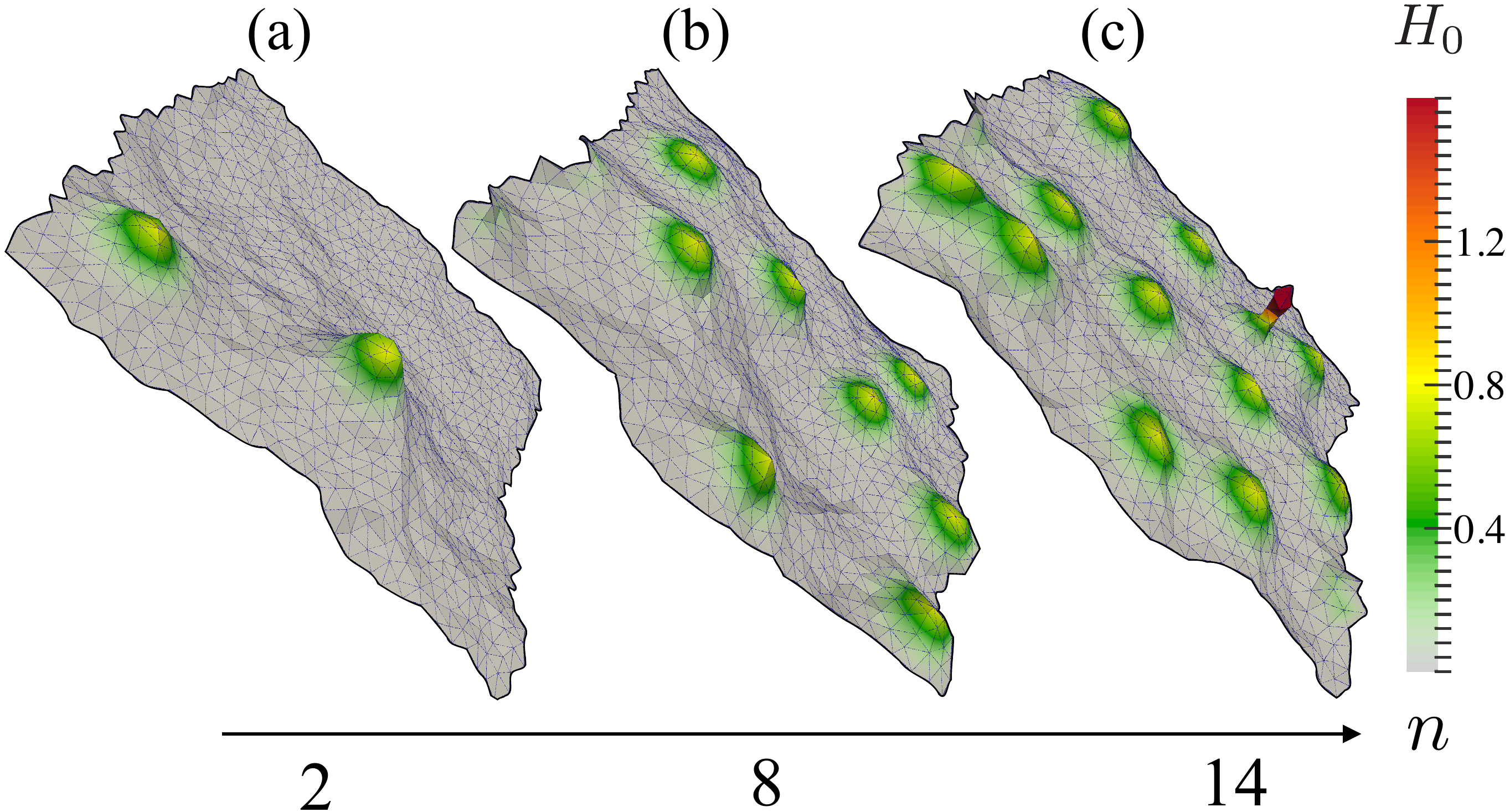}
\caption{\label{fig:czero-0p8-funcepsin}{\small \sl(Color Online)} Representative membrane conformations as a function of epsin concentration for $C_0 = 0.8 a_0^{-1}$: (a) 2 protein fields; (b) 8 protein fields; (c) 14 protein fields. Color bar shows the induced curvature field $H_{0}$ in units of $a_0^{-1}$; a tubule is present in (c).}
\end{figure}

With increase in protein concentration, spatial aggregation is more pronounced and cooperative effects\textemdash due to membrane curvature-mediated interactions\textemdash stabilize protein clustering as well as induce morphological transitions. In this limit, the proteins collectively induce stable morphological features in the membrane as seen in Figure~\ref{fig:czero-0p8-funcepsin}; here, for $C_0=0.8a_0^{-1}$, protein clustering leads to tubule formation when $n>10$. 

Figures \ref{fig:epsin6-funcczero} and \ref{fig:czero-0p8-funcepsin} show  how  the control variables such as $A/A_P$ (relative membrane area) and $n$ (protein concentration) govern the emergent membrane morphologies. These  results also suggest subtle competition between the translational entropy of proteins, entropy due to membrane undulations, and membrane deformation energy due to curvature-induction by proteins. Since  the morphological changes in the membrane are associated with a large change in the entropy of the system, they can be quantitatively tracked only by computing the  free-energy landscape of curvature induction.  The following sections quantify the free-energy landscape of protein-induced curvature deformations as a function of $C_0$ and $n$.  


\subsection{Widom Test Particle/Field Insertion Method}
\begin{figure}[!h]
\centering
\includegraphics[width=7.5cm]{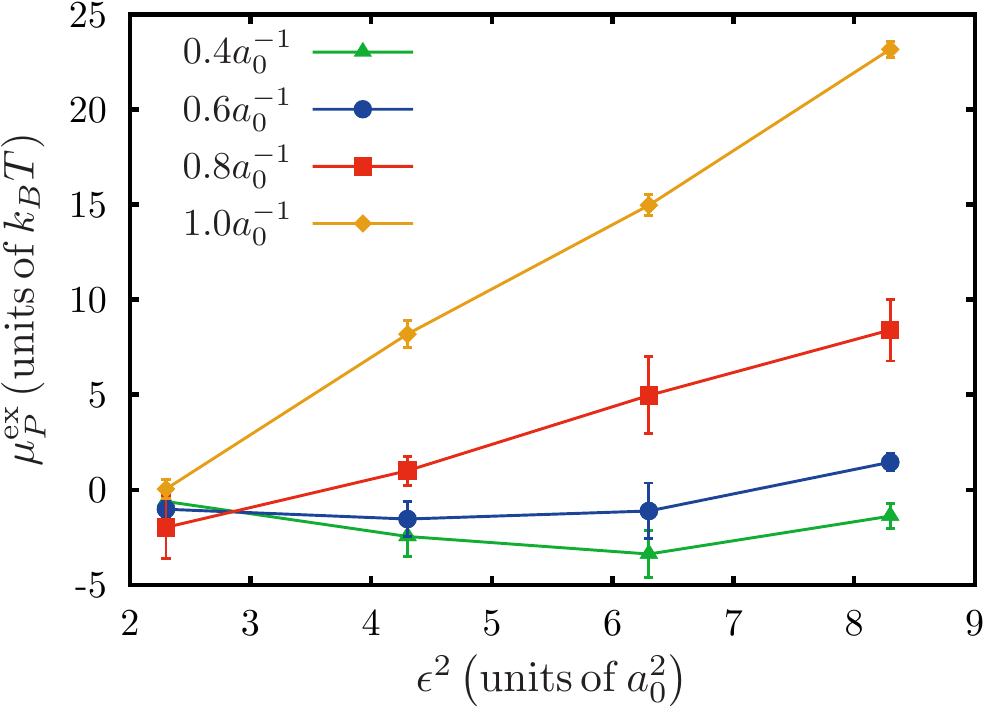}  
\caption{{\small \sl(Color Online)} Excess chemical potential as a function of $\epsilon^2$, for an isotropic Gaussian curvature obtained through the Widom test-particle/field insertion method.  Data shown for four values of spontaneous curvature, $C_0$.\label{fig:planarhomogeneousc0}}  
\end{figure}

Widom test-particle/field insertion method is used to quantify the excess chemical potential of curvature-inducing proteins on a planar membrane. Figure~\ref{fig:planarhomogeneousc0} shows the excess chemical potential for dilute protein concentrations (i.e., $n \rightarrow 0$)  as a function of $C_0$ and $\epsilon^2$, for curvature fields of the form given by equation~\eqref{eqn:curiso}.  For $C_0 = 0.4 a_0^{-1}$ and $0.6 a_0^{-1}$, $\mu_{P}^{\rm ex}$ is negative, and hence it is favorable to insert a protein on the membrane.  In this limit, the protein's curvature field is shallow and matches well with the equilibrium curvature profile of the natural undulations in the membrane leading to reduced free-energy/chemical potential.  However, it should be noted that the excess chemical potential can cross over to positive values with further increase in the value of  $\epsilon^2$ and the insertion of a protein is no longer thermodynamically favorable. For $C_0 = 0.8 a_0^{-1}$ and $1.0 a_0^{-1}$, the crossover to positive $\mu_{P}^{\rm ex}$ is observed at much lower values of $\epsilon^2$.  $\mu_{P}^{\rm ex}$ increases linearly with $\epsilon^2$ with their respective slope depending on the value of $C_0$. An increase in $\mu_{P}^{\rm ex}$ is a signature of curvature induced deformation, since equilibrium membrane profiles cannot accommodate such large curvatures. Hence, these results quantify both the {\it curvature-sensing} and {\it curvature-inducing} behavior of membrane proteins. 

\begin{figure}[!h]
\centering
\includegraphics[width=7.5cm]{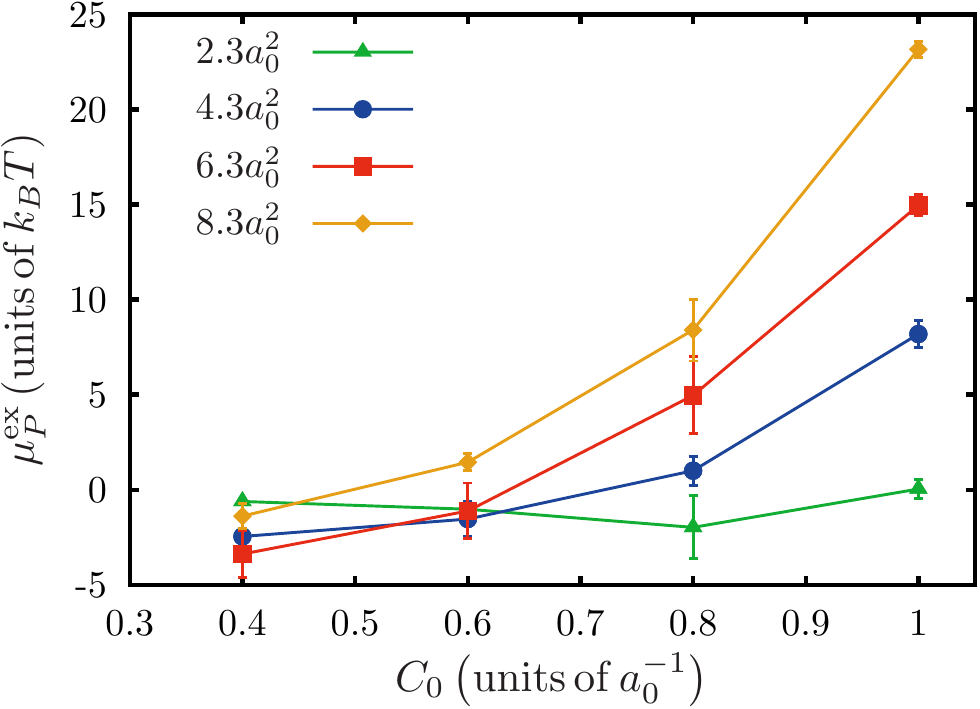} 
\caption{{\small \sl(Color Online)} Excess chemical potential as a function of $C_0$, for an isotropic Gaussian curvature obtained through the Widom test-particle/field insertion method.  Data shown for four values of variance, $\epsilon^2$. \label{fig:planarhomogeneousr2}}  
\end{figure}

The excess chemical potential as a function of the induced spontaneous curvature $C_0$  is shown in Figure~\ref{fig:planarhomogeneousr2} (data from Figure~\ref{fig:planarhomogeneousc0} has been replotted).  As stated before, the free-energy for insertion of a protein is negative for small magnitudes of  induced curvature and extents (low $C_0$ and $\epsilon^2$).  For higher values of $\epsilon^2$ the excess chemical potential is observed to grow quadratically with $C_0$, as predicted by equation~\eqref{eqn:analytic-excess-simplified}. We note that the higher values of $\epsilon^2$ correspond to an energy dominated regime, for which, by relative comparison, the entropic correction (second term in RHS of equation~\eqref{eqn:analytic-excess-simplified}) is small enough to be neglected.

\begin{figure}[!h]
\centering
\includegraphics[width=7.5cm]{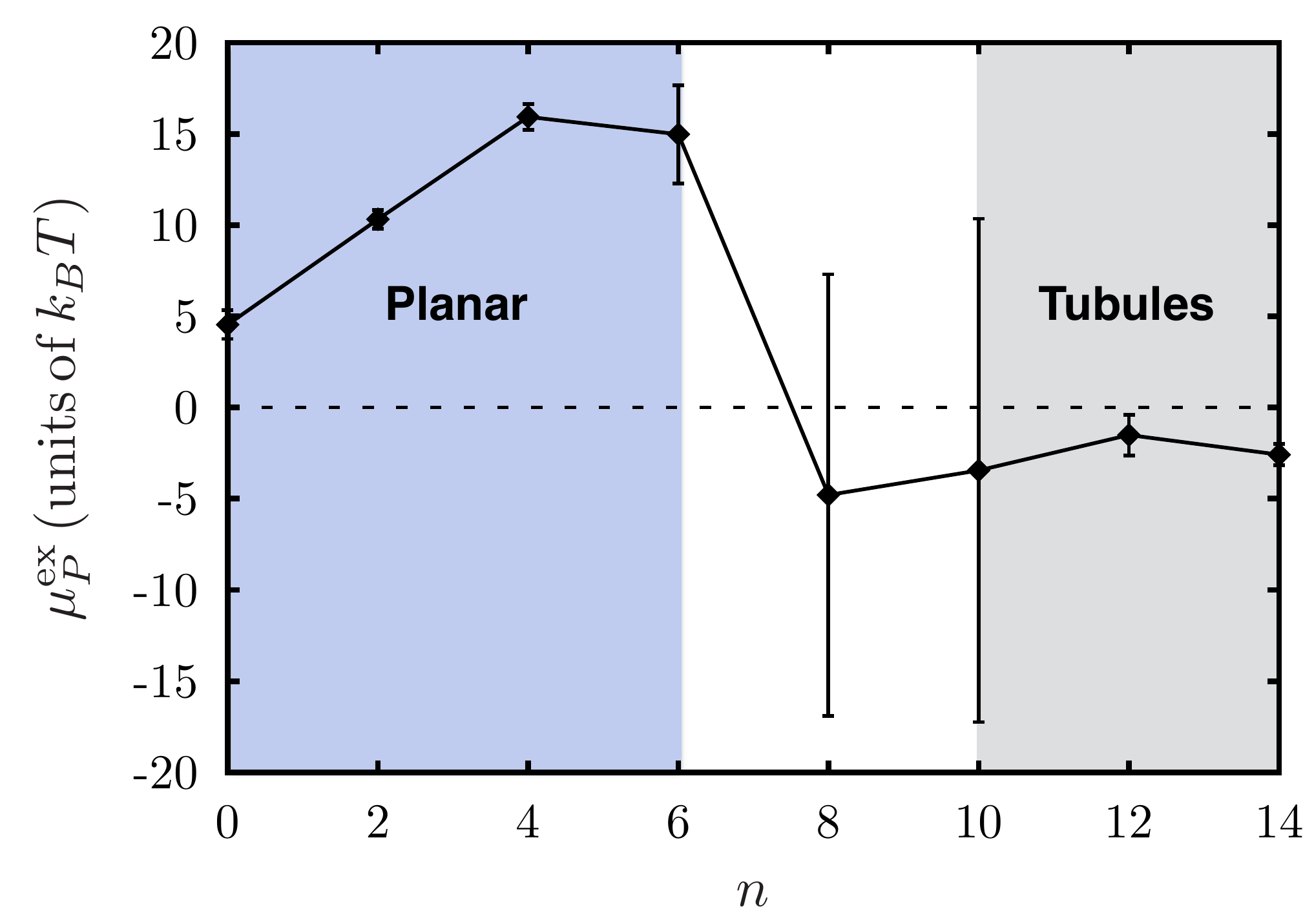} 
\caption{{\small \sl(Color Online)} Excess chemical potential of an isotropic Gaussian curvature field, with $C_0 = 0.8 a_0^{-1}$ and $\epsilon^2 = 6.3 a_0^2$, as a function of the number of proteins $(n)$. \label{fig:planarhomogeneousconc}}  
\end{figure}

We have shown in Figure~\ref{fig:planarhomogeneousconc} the computed chemical potential as a function of protein concentration $(n)$ for a planar membrane with $C_0 = 0.8 a_0^{-1}$ and $\epsilon^2 = 6.3 a_0^2$.  For small values of $n$ where the concentration of proteins does not considerably affect the membrane undulation, we observe $\mu_{P}^{\rm ex}$ to be positive and to increase with increasing value of $n$.  The excess chemical potential reaches a peak value at $n\approx6$\textemdash beyond which the chemical potential drops to negative values implying that the subsequent recruitment of proteins is favorable.  In analogy, the region to the left of the peak corresponds to the planar membrane morphology shown in Figure~\ref{fig:czero-0p8-funcepsin}(a) and the region marked tubules to the extreme right corresponds to the tubulated membrane conformation shown in Figure~\ref{fig:czero-0p8-funcepsin}(c).  In the transition region we observe both  tubulated and planar morphologies with equal probabilities. This leads to large fluctuations  in $\mu_{P}^{\rm ex}$,  which is indicated by the large error bars in the chemical potential for protein concentrations $n=8$ and $n=10$ (see Figure~\ref{fig:planarhomogeneousconc}).
The above example demonstrates that the Widom test-particle/field method is a powerful approach to quantitatively map the phase boundary associated with morphological transitions in membranes.



\subsection{Comparison to Analytical Results}
The values of the chemical potential at infinite dilution can also be computed analytically.  However, for proteins with finite curvature extent, a direct comparison with analytical results is complicated by the non-trivial curvature-field dependent term in  equation~\eqref{eqn:analytic-excess-simplified}.  It is possible, however, to obtain closed form analytical predictions for the excess chemical potential when spontaneous curvature fields of the form $H_0 = C_0 \delta\left(r-r'\right)$ are considered.  In this section, the results obtained from the Widom test-field method are compared against analytical predictions for such curvature fields.

\begin{figure}[!h]
\centering
\includegraphics[width=7.5cm]{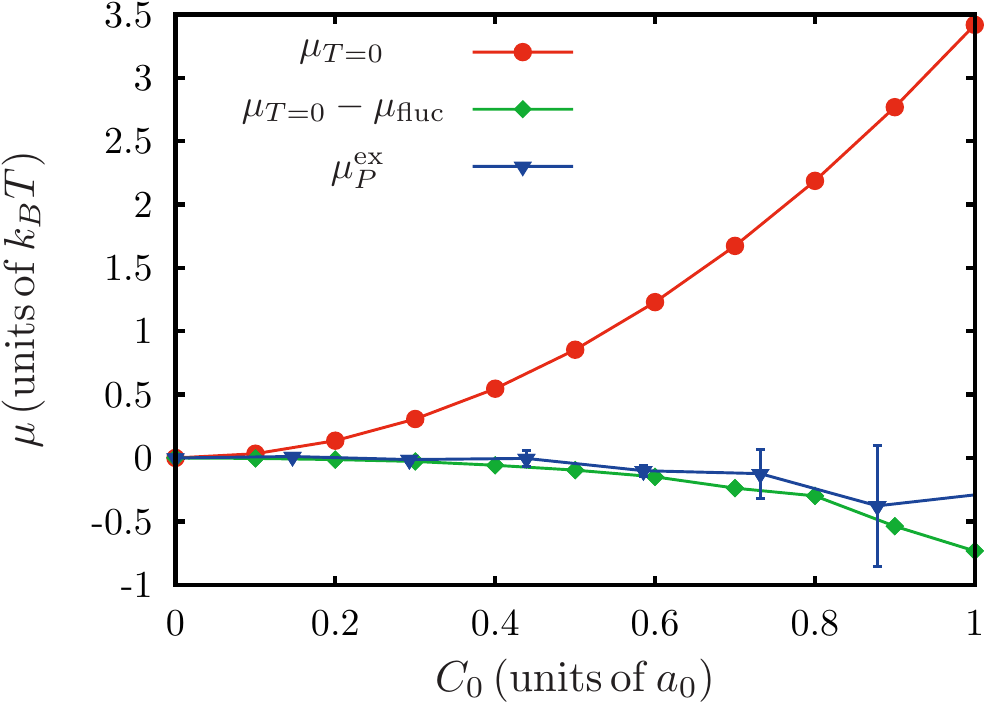} 
\caption{{\small \sl(Color Online)} Widom test particle/field insertion results are shown in blue. Analytical scaling with a fluctuation correction calculated from the cumulant expansion is shown in green. Zero temperature scaling shown in red. \label{fig:compare}}  
\end{figure}

In our model, proteins which do not have large extents of curvature can be approximated as point sources of spontaneous curvature.  A point spontaneous curvature field can be described by, 
 \begin{equation}
 H_0(\vec{r}_{m},\vec{r}_{p}) =  C_0 \delta(r), \, \textrm{where} \, r=|\vec{r}_{m}-\vec{r}_{p}|.
\label{eqn:dirac-c0}
 \end{equation}
 Using equation~\eqref{eqn:dirac-c0}, equation~\eqref{eqn:analytic-excess} can be reduced to,
\begin{eqnarray}
\mu_{P}^{\rm ex} & = &\underbrace{ \frac{\kappa C_0^2}{2 A_{\rm vertex}}}_{\mu_{T=0}} - \underbrace{{k_B T \ln\left\langle \exp\left( \frac{2 \kappa C_0}{k_B T} H(s_{n+1}) \right)\right\rangle}_n}_{\mu_{\rm fluc}}. 
\label{eqn:dirac-uexc}
\end{eqnarray}
Here, $A_{\rm vertex}=\sqrt{3}(1.3 a_0)^2/2$ is the area per vertex in our discrete triangulated mesh, and $1.3 a_0$ is the average link length at the value of $A/A_{P}$ employed here: the factor $A_{\rm vertex}$ arises because of the discrete approximation to the Dirac delta function. The ensemble average in \eqref{eqn:dirac-uexc} can be evaluated in simulations through a cumulant expansion, 
\begin{equation}
\langle \exp\left( tH \right) \rangle = 1 + t \langle H^1 \rangle + \frac{t^2}{2!} \langle H^2 \rangle + \frac{t^3}{3!} \langle H^3 \rangle + ...,
\label{eqn:cumulant-eqn}
\end{equation}
where, $\langle H^i \rangle$ is the $i$'th moment of the mean curvature, and $t=2\kappa C_0/k_B T$.  As demonstrated in Appendix~\ref{app:cumulant}, the sum of terms $\langle H^i \rangle$ is a weakly decaying function of $i$, and hence we retain the first 15 terms in order to obtain convergence. In Figure~\ref{fig:compare}  $\mu_P^{\rm ex}$ obtained from the Widom test-field method is plotted and compared against $\mu_{T=0}$ and  $\left(\mu_{T=0}-\mu_{\rm fluc}\right)$.   The analytical results with finite temperature corrections agree well with $\mu_{P}^{\rm ex}$. The Widom test-field method is thus validated for point spontaneous curvature fields, and hence we are confident that the method gives reliable estimates for the excess chemical potential. It should be noted that the fluctuation corrections ($\mu_{\rm fluc}$) for the point spontaneous curvature field ranges from $0$ to $6 k_B T$.  This large correction is a manifestation of the protein curvature field localizing to membrane undulations matching their profile, and the value of $\mu_{\rm fluc}$ depends on $\kappa$, $C_0$, $\epsilon^2$, and $n$. In the next section, results from the Widom test-particle/field insertion method is compared to results from both thermodynamic integration and Bennett acceptance methods, to further validate the estimates for the chemical potential.


\subsection{Comparison of Free-Energy Methods}
The chemical potential to insert a protein field is given by, 
\begin{equation}
\mu =  \frac{dF}{dn} = \frac{\Delta F}{\Delta n}.
\end{equation}
Here, $\Delta F$ is the free-energy change to insert $\Delta n$ proteins.  We compute $\Delta F$ using two free-energy perturbation techniques namely thermodynamic integration and Bennett acceptance method (BAM).  The techniques used to compute $\Delta F_{\rm TI}$ and $\Delta F_{\rm BAM}$ involve growing $\Delta n$ curvature fields that have zero spontaneous curvature initially (state $A$), to a desired value of $C_0$ (state $B$).  Since the presence of the protein is felt only through $C_0$, perturbing the system from state $A$ to $B$ is analogous to inserting $\Delta n$ proteins.

\begin{figure}[!h]
\centering
\includegraphics[width=7.5cm]{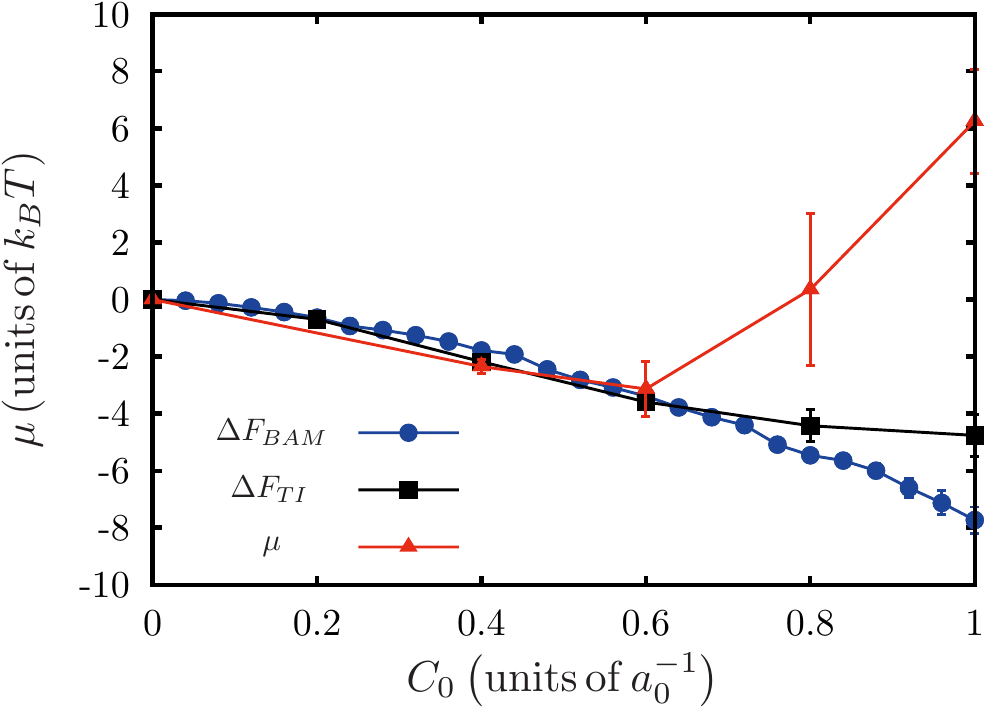} 
\caption{{\small \sl(Color Online)} Comparison of chemical potentials from Widom, TI, and BAM, for different $C_0$ with $\epsilon^2= 6.3 a_0^2$ and $n=1$. \label{fig:method-comparison}}  
\end{figure}

In order to make direct comparisons to results from the Widom test-field method, we choose $\Delta n = 1$. In this case, the values of $\Delta F_{\rm TI}$ and $\Delta F_{\rm BAM}$ are related to the chemical potential as,
 \begin{equation}
\Delta F_{\rm TI} = \Delta F_{\rm BAM} = \mu = \mu_P^{\rm id} (\rho) + \mu_P^{\rm ex},
\end{equation}
where, $\mu_P^{\rm ex}$ is calculated using Widom test particle/field insertion method, while the configurational contribution  to the entropic correction, $\mu_P^{\rm id} (\rho)$ (as discussed above in eqn.  \eqref{eqn:widom}), is given by,
\begin{equation}
\mu_P^{\rm id}(\rho) = k_B T \ln \rho.
\end{equation}
Both thermodynamic integration and Bennett acceptance methods calculate the difference in free-energy between a state with no protein field and a state with one protein field.  Results from Widom insertion method cannot be directly compared to TI or BAM to an important difference in sampling between the methods.  Widom insertion samples the curvature field equally at all spatial locations on the membrane, whereas TI and BAM introduce the curvature field at a specific spatial location; this difference in sampling defines a correction of entropic origin for thermodynamic integration and Bennett methods which needs to be accounted for before all three methods can be compared against one another. Details of the procedure for computing $\mu_P^{\rm id}(\rho)$ are given in Appendix~\ref{app:entropic-correction}.  The values of $\Delta F_{\rm TI}$ and $\Delta F_{\rm BAM}$ are plotted and compared against Widom test-field values for $\mu$ in Figure~\ref{fig:method-comparison}.  The results show excellent agreement for small values of $C_0$, but each method deviates as the spontaneous curvature is increased.  The estimate for the chemical potential $\mu$ agrees very well with $\Delta F_{\rm TI}$ and $\Delta F_{\rm BAM}$ for small values of $C_0 < 0.6 a_0^{-1}$.  For larger values of $C_0$ the chemical potential determined using the Widom method deviates from the estimates derived from the perturbation techniques.  The comparison between the methods at higher protein densities is also investigated and the results are discussed in Appendix~\ref{app:high-densities}.

The mismatch in the values of $\mu$ between these methods at large values of $C_0$ is well known.  In the case of Widom test-field insertion the deviation is a result of dominant contributions from some rare conformations to the chemical potential. Estimates for the chemical potential from thermodynamic integration also break down due to insufficient sampling at larger values of $C_0$; this can be seen in the small deviations from the corresponding values of BAM in Figure~\ref{fig:method-comparison}.  Metrics to quantify the sampling error from the three methods are discussed further in Appendices~\ref{app:widom}, \ref{app:ti}, and ~\ref{app:bennett}. The applicability of each of the described free-energy methods depends on the system investigated, desired accuracy, and the available computational resources.  The Widom insertion technique gives accurate results with low computational overhead and this works very well for dilute protein concentrations and weak curvature fields; however at higher protein concentrations and strong curvature fields, where the energies are large, this method becomes inaccurate and this is a know artifact of Widom insertion.  On the other hand perturbative techniques like TI and Bennett work very well for all concentrations, but are computational expensive.  For dense systems TI or Bennett methods are better suited.
\\

\section{Conclusion} 
Three free-energy sampling methods have been used to quantify the chemical potential of curvature-inducing proteins in a field theoretic mesoscale cell membrane model.  Results show good agreement between each method for weak spontaneous curvature fields and deviate at strong curvature field strengths due to the differences in the nature of sampling in each method.  The results from the Widom method are also in excellent agreement with an analytical result for curvature fields approximated by a delta function, further validating our computational approach. The analytical result also provides a basis to explain the quadratic dependence of the excess chemical potential on the strength of the curvature field induction in an energy dominated regime. Further, the utility of the Widom particle/field insertion method to quantitatively track phase boundaries associated with morphological transitions of the membrane has been successfully demonstrated in the context of a tubulation transition. Our results also indicate that the Widom test particle/field insertion method fails to capture the correct chemical potential at high curvature field strengths, as expected, due to the large perturbation in energy. In this limit, the thermodynamic integration and the Bennett acceptance methods perform favorably to control the statistical error. With these caveats noted, the free-energy approach to quantify the energy landscape of protein-mediated membrane deformation is novel and powerful in quantitatively examining protein-induced morphological transitions in bilayer and membrane systems. Our simulations are able to recapitulate a tubulation transition above a critical density of curvature-inducing proteins.  Tubulation of liposomes has been widely observed in the literature for high concentrations of curvature-inducing proteins including Epsin, Amphiphysins, and Exo70 \cite{Ford:2002if,Masuda:2006kc,Zhao:2013hi}.  Given the characteristics of a single protein curvature field, our model would be able to predict these tubulation thresholds for each protein species. Future work will focus on extending these methods to study curvature-sensing in cylindrical/tether geometries, anisotropic curvature fields, systems with inhomogeneous background curvature, and also the effect of control variables such as tension in morphological transitions of the membrane induced by spontaneous curvature. 

\section*{Acknowledgments}
This work was supported in part by National Science Foundation grants DMR-1120901, CBET-1133267, and CBET-1244507, and the National Institutes of Health grant NIH U01-EB016027. The research leading to these results has received funding from the European Commission grant FP7-ICT-2011-9-600841. Computational resources were provided in part by the National Partnership for Advanced Computational Infrastructure under Grant No. MCB060006 from XSEDE. 

\appendix

\section{Widom Test Particle/Field Insertion: Quantification of Sampling} \label{app:widom}
\begin{figure}[!h]
\centering
\includegraphics[width=7.5cm]{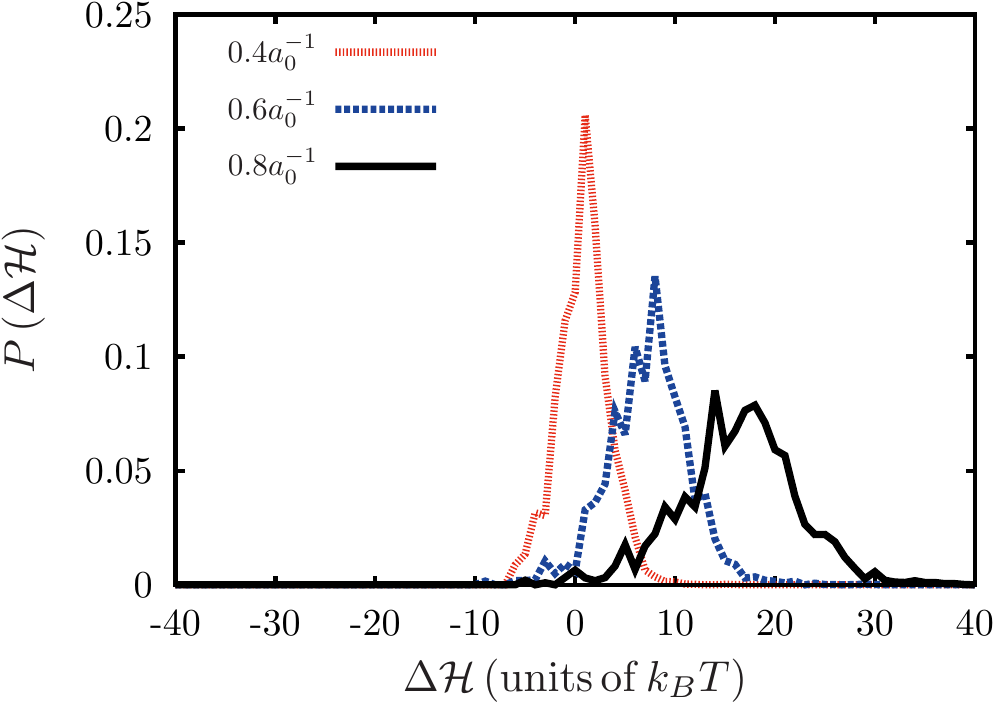}  
\caption{{\small \sl(Color Online)} Normalized distribution of $\Delta \cal{H}$ obtained using the Widom particle/field insertion method for several $C_0$; here $\epsilon^2 = 6.3 a_0^2$. \label{fig:widom-histogram}}  
\end{figure}

In the Widom test particle/field insertion method, the ensemble average is taken over a Boltzmann distribution of $\Delta \cal{H}$. This means, the small or negative $\Delta \cal{H}$ values will dominate the ensemble average. The distribution of $\Delta \cal{H}$ is a Gaussian, as shown in Figure~\ref{fig:widom-histogram}, with $P(\Delta \cal{H})$ dependent on the strength of the curvature field. As the strength of the curvature field increases the mean of this Gaussian distribution will shift to the right, towards higher energies and both the precision and accuracy of the Widom method will be impacted adversely.
%
\section{Accuracy of Thermodynamic Integration}
\label{app:ti}
\begin{figure}[!h]
\centering
\includegraphics[width=7.5cm]{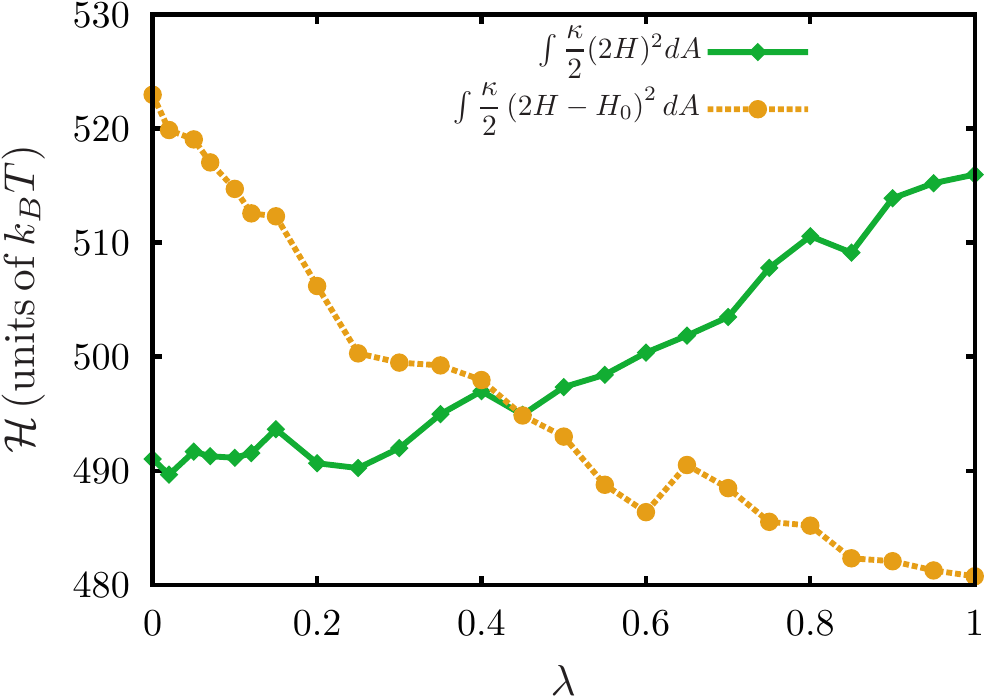} 
\caption{{\small \sl(Color Online)} Plots of ${\cal{H}}_{\lambda=0}$ and ${\cal{H}}_{\lambda=1}$ as a function of $\lambda$. Data shown corresponds to a spontaneous curvature field with $C_0=0.8 a_0^{-1}$ and $\epsilon^2=6.3 a_0^2$.} 
\label{fig:ti-e-e0}
\end{figure}

By setting up a range of simulations over the Kirkwood coupling parameter, $\lambda$, in the interval from 0 to 1, the elastic energy of the membrane with and without $H_0$ can be tracked and integrated along $\lambda$.  Figure~\ref{fig:ti-e-e0} details the contributions of ${\cal{H}}_{\lambda=0}$ and ${\cal{H}}_{\lambda = 1}$.  To calculate the chemical potential, which can be compared to Widom insertion, the free-energy is computed by introducing one spontaneous curvature field $\left( \Delta n = 1 \right)$.
%
\section{Accuracy of Bennett Acceptance Method}
\label{app:bennett}
\begin{figure} [ht]
\centering
\includegraphics[width=7.5cm]{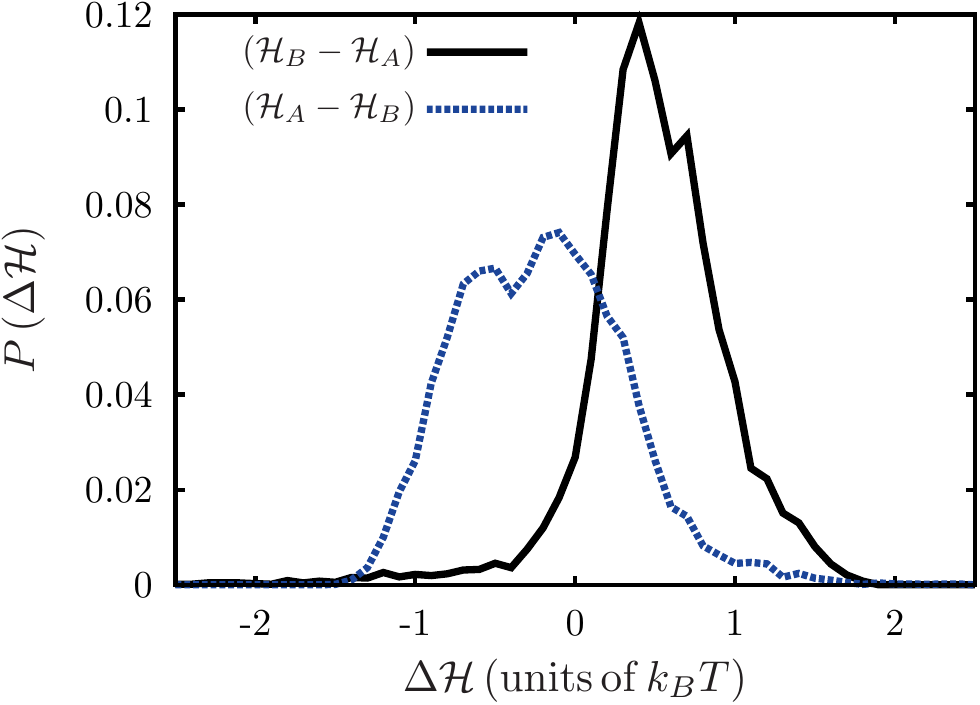}
\caption{{\small \sl(Color Online)} Normalized distribution of the change in energy when the membrane transits from state $A$ to state $B$ and vice-versa in BAM. \label{fig:bennett-histogram}}  
\end{figure}

The Bennett Acceptance method requires the two states being sampled to have a small difference in energy. This accuracy can be quantified by plotting the distribution of $\Delta {\cal H}$ in each direction sampled ($A \rightarrow B$ and $B \rightarrow A$). A large overlap in the distributions of $\Delta \cal{H}$ describes states which have a small difference in energy.  For example, in the case of one curvature-inducing protein, the states $A$ and $B$ represent a membrane with curvature fields $C_0$ and $C_0 + \delta C_0$, respectively. Consider state $A$ to have a spontaneous curvature $C_0 = 0.8 a_0^{-1}$, and state $B$ to have $C_0 = 0.76 a_0^{-1}$, for a fixed $\epsilon^2 = 6.3 a_0^2 $.  The normalized distribution of $\Delta \cal{H}$ is shown in Figure~\ref{fig:bennett-histogram}.  As expected, the energy is normally distributed, with the overlap between each distribution being within one standard deviation of each other.  If the states are separated further apart in energy, this overlap will become minimal, and the accuracy of Bennett will decline.

\section{Convergence of the Cumulant Expansion}
\label{app:cumulant}
\begin{figure}[!h]
\centering
\includegraphics[width=7.5cm]{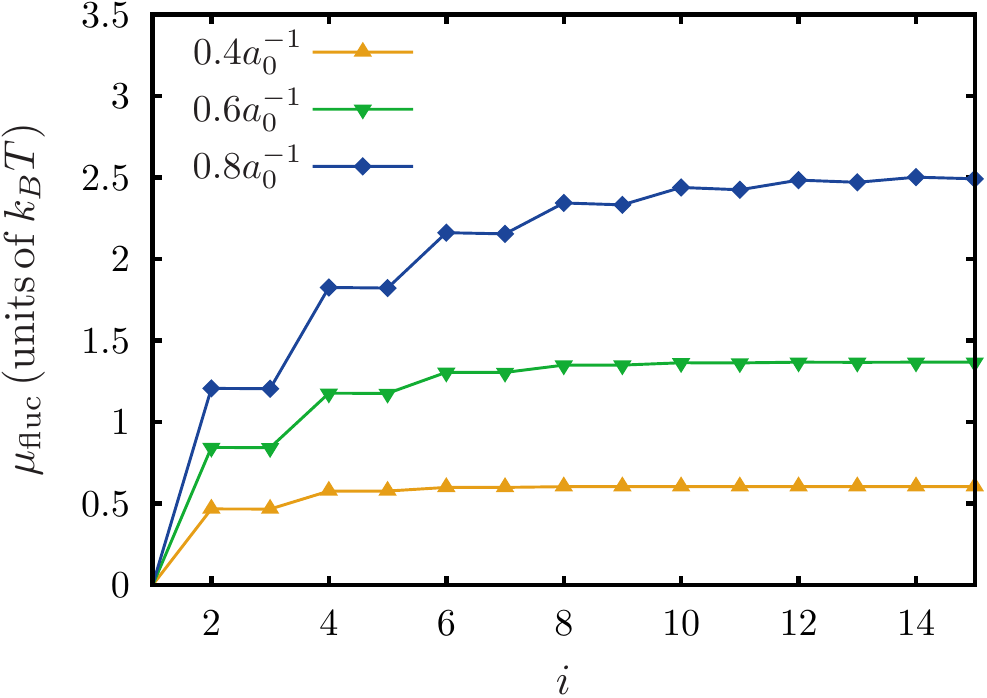} 
\caption{{\small \sl(Color Online)} The fluctuation correction for the chemical potential ($\mu_{\rm fluc}$) obtained with the cumulant expansion as a function of the number of terms $(i)$ considered.  Data shown for three values of $C_0 = 0.4,~0.6,~0.8~a_0^{-1}$. \label{fig:converge}}  
\end{figure}

The number of terms to be retained in a cumulant expansion depends upon its convergence behavior. Figure~\ref{fig:converge} shows $\mu_{\rm fluc}$ computed using a cumulant expansion as a function of the number of terms ($i$) retained.  It can be seen that for higher $C_0$ more terms need to be considered in order to attain convergence.  For all analysis presented in this article, the first 15 terms were used to compute $\mu_{\rm fluc}$.

\section{Estimation of the Entropic Correction} \label{app:entropic-correction}
In order to compare TI or Bennett with Widom insertion method, the difference in density sampling between the methods can be approximated. In TI or BAM the spontaneous curvature field stabilizes a bump on the membrane and this limits the lateral diffusion of membrane protein field.  This means that in the limit of a large $C_0$ and $\epsilon^2$, the membrane curvature field can only sample a small region of the membrane which cuts off entropic contributions due to diffusion.  In a Widom simulation the curvature field probes the free-energy with equal probability across the whole membrane.  This disparity in density sampling is of entropic origin and can be written as,
\begin{equation}
F_{\rm TI/BAM} + F \left( \rho \right) = \mu_P^{\rm ex}
\end{equation}
where 
\begin{equation}
F \left( \rho \right) = - k_B T \ln \left( \frac{2 \sigma_\psi}{N_{\rm vert}} \right),
\end{equation}
with $\sigma_\psi$ being some average number of vertices out of a total $N_{\rm vert}$ vertices that a curvature field visits in a TI or Bennett simulation.  The entropy lost in a thermodynamic integration simulation was computed by plotting a histogram of the number of unique vertices visited by an curvature field, $\psi$, and finding the standard deviation of that distribution, $\sigma_\psi$.  The standard deviation is calculated from
\begin{equation}
\sigma_{\psi}^2 = \sum_j j^2 P \left( j \right)^2 - \sum_j \left( j P\left( j \right) \right)^2.
\end{equation}
Calculated values of standard deviation and their corresponding values of free-energy are listed in Table~\ref{table:unique-vert}.
\begin{figure}[ht]
\centering
\includegraphics[width=7.5cm]{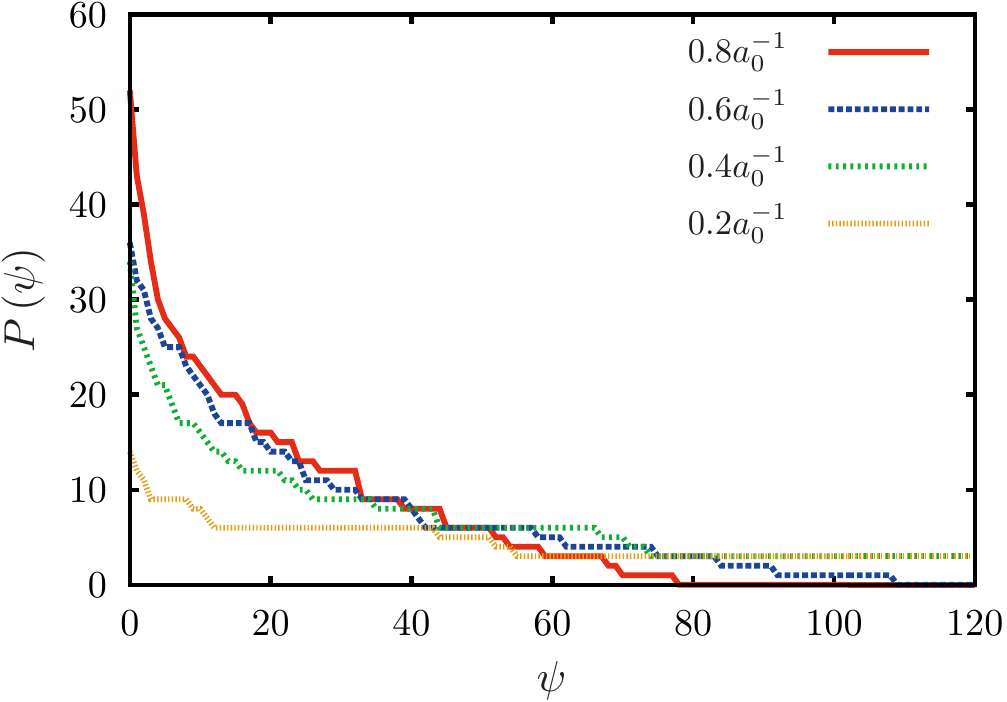} 
\caption{{\small \sl(Color Online)} Histogram of the number of unique vertices visited by a curvature field $\left( \psi \right)$ in a TI simulation as a function of $C_0$. \label{fig:unique-vert-dist}}  
\end{figure} 
\begin{table}[ht]
\caption{Estimation of the Entropic Correction}
\centering
\begin{tabular}{c c c}
\hline\hline
$C_0/(a_0^{-1})$& ~~~$2 \sigma_{\psi}$~~~ & $F \left( \rho \right) / \left( k_B T \right)$  \\ [0.5ex]
\hline 
0.2 &   242 & 1.31 \\ [0.4ex]
0.4 &   105 & 2.15 \\ [0.4ex]
0.6 &   74 & 2.50 \\ [0.4ex]
0.8 &   49 & 2.90 \\ [0.4ex]
1.0 &   37 & 3.17 \\ [1.0ex]
\hline
\end{tabular}
\label{table:unique-vert}
\end{table}

\section{Comparison of Free-energy Methods at Higher Densities} \label{app:high-densities}
\begin{figure}[ht] 
\centering
\includegraphics[width=7.5cm]{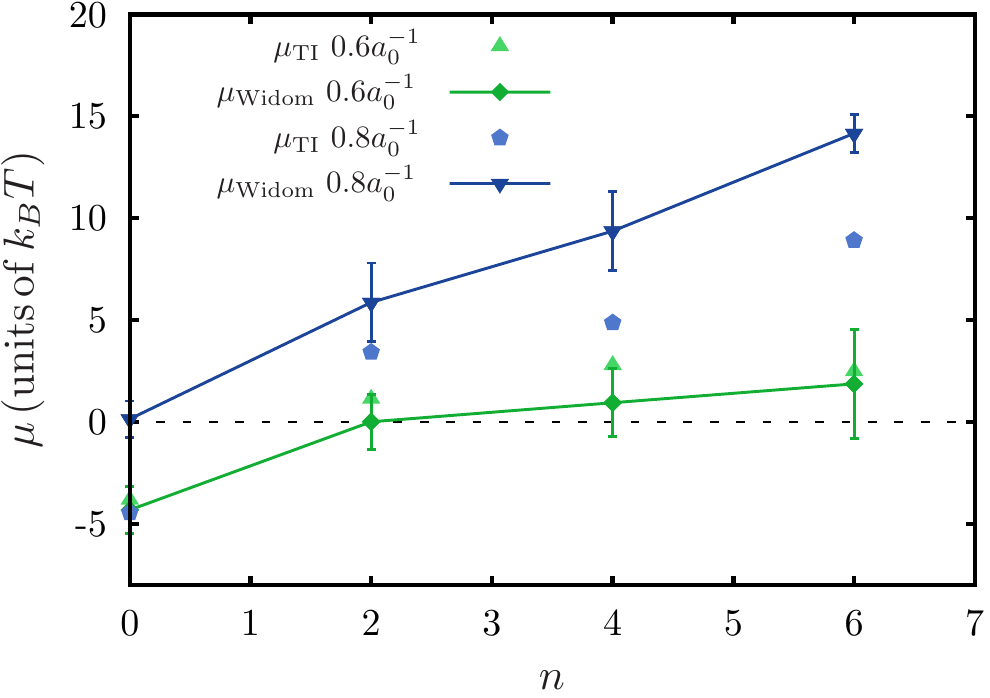}  
\caption{{\small \sl(Color Online)} Chemical potential versus $n$:  Data from the results of the Widon method and the TI method are shown for two $C_0$s with $\kappa = 10 k_B T$ and $\epsilon^2 = 6.3 a_0^2$. \label{fig:ti-widom-conc}}  
\end{figure}

The Widom particle/field insertion method is known to fail at high densities due to the nature of its sampling. Therefore a comparison of free-energy methods for higher densities is done in order to quantify its accuracy.  A comparison between the chemical potential obtained from both TI and the Widom method for several protein concentrations ranging from $n=0$ to $n=6$ is shown in Figure~\ref{fig:ti-widom-conc}.  The entropic correction for the Widom method is calculated according to Appendix~\ref{app:entropic-correction}. For $C_0 = 0.8 {a_0}^{-1}$ this correction is approximately $F(\rho) = 2.85 k_B T$, for $C_0 = 0.6 {a_0}^{-1}$ its $F(\rho) = 2.42 k_B T$, and for $C_0 = 0.4 {a_0}^{-1}$ its $F(\rho) = 1.83 k_B T$.  The comparison in Figure~\ref{fig:ti-widom-conc} shows that the methods agree within statistical error for $C_0=0.6 {a_0}^{-1}$.  The deviation between the results at $C_0 = 0.8 {a_0}^{-1}$ is systematic and is expected due to a similar deviation seen in Figure~\ref{fig:method-comparison} between the Widom method and other free-energy methods for dilute concentrations as discussed in Figure~\ref{fig:method-comparison}.

\bibliography{bibfile-9May14}

\end{document}